\documentclass[aps,prl,reprint,superscriptaddress,amsmath,amssymb,showpacs]{revtex4-1}

\usepackage{graphicx}
\usepackage{color}
\usepackage{bbm}

\date{\today}                  

\begin{document}

\title{Chiral 1D Floquet topological insulators beyond rotating wave approximation}

\author{Dante M. Kennes}
\affiliation{Dahlem Center for Complex Quantum Systems and Fachbereich Physik, 
Freie Universit\"at Berlin, 14195 Berlin, Germany}
\author{Niclas M\"uller}
\affiliation{Institut f\"ur Theorie der Statistischen Physik, RWTH Aachen, 
52056 Aachen, Germany and JARA - Fundamentals of Future Information Technology}
\author{Mikhail Pletyukhov}
\affiliation{Institut f\"ur Theorie der Statistischen Physik, RWTH Aachen, 
52056 Aachen, Germany and JARA - Fundamentals of Future Information Technology}
\author{Clara Weber}
\affiliation{Institut f\"ur Theorie der Statistischen Physik, RWTH Aachen, 
52056 Aachen, Germany and JARA - Fundamentals of Future Information Technology}
\author{Christoph Bruder}
\affiliation{Department of Physics, University of Basel, Klingelbergstrasse 82, CH-4056 Basel, Switzerland}
\author{Fabian Hassler}
\affiliation{JARA-Institute Quantum Information, RWTH Aachen University, 52056 Aachen, Germany}
\author{Jelena Klinovaja}
\affiliation{Department of Physics, University of Basel, Klingelbergstrasse 82, CH-4056 Basel, Switzerland}
\author{Daniel Loss}
\affiliation{Department of Physics, University of Basel, Klingelbergstrasse 82, CH-4056 Basel, Switzerland}
\author{Herbert Schoeller}
\email[Email: ]{schoeller@physik.rwth-aachen.de}
\affiliation{Institut f\"ur Theorie der Statistischen Physik, RWTH Aachen, 
52056 Aachen, Germany and JARA - Fundamentals of Future Information Technology}

\begin{abstract}
We study one-dimensional (1D) Floquet topological insulators with chiral symmetry  
going beyond the standard rotating wave approximation. The occurrence of many
anticrossings between Floquet replicas leads to a dramatic extension of 
phase diagram regions with stable topological edge states (TESs). We present an explicit  
construction of all TESs in terms of a truncated Floquet Hamiltonian in frequency space, prove
the bulk-boundary correspondence, and analyze the stability of the TESs in 
terms of their localization lengths. We propose experimental tests of our predictions 
in curved bilayer graphene. 

\end{abstract}


\maketitle

The search for materials in which topologically protected edge states can be induced and controlled 
has led to much research attention in the field of Floquet topological insulators (FTIs) 
in the last decade. 
In FTIs topology is induced by external driving which implies an intriguing degree of tunability.
These systems were classified [\onlinecite{kitagawa_etal_10}-\onlinecite{hoeckendorf_etal_prb_18}] and first 
experimental tests were discussed in photonic crystals
[\onlinecite{kitagawa_etal_natcomm_12}-\onlinecite{mukherjee_etal_natcomm_17}],
cold atom systems [\onlinecite{jotzu_etal_nature_14}-\onlinecite{quelle_etal_njp_17}],
and solid state materials [\onlinecite{calvo_etal_apl_11}-\onlinecite{wang_scirep_17}]. 
Many promising proposals were made for realizing, e.g.~, Majorana edge modes 
\cite{jiang_etal_prl_11,kundu_seradjeh_prl_13,thakurathi_etal_prb_13,thakurathi_sengupta_sen_prb_14,
reynoso_frustaglia_prb_13} and parafermions \cite{thakurathi_loss_klinovaja_prb_17} in 1D FTIs, 
the photo-induced quantum
Hall effect in 2D materials [\onlinecite{oka_aoki_prb_09}-\onlinecite{perez-piskunow_etal_pra_15}],
topological surface states in 3D FTIs \cite{lindner_etal_prb_13}, and Weyl semimetals
and fractional FTIs in coupled Rashba nanowires \cite{klinovaja_stano_loss_prl_16}.

A particular feature of FTIs distinguishing them fundamentally from static TIs is the fact that the driving
frequency $\omega$ leads to an infinite set of Floquet replicas defined by shifting the conduction and 
valence band by $n\omega$ \cite{hbar} corresponding to absorption/emission of $n$ photons. This leads to
a hierarchy of Floquet-induced anticrossings in the center of the bulk gap, which occurs  
for all solid state realizations of FTIs where the band width $W$ is typically large compared to $\omega$. 
This property has received only little attention so far. 
Previous works considered either the case of large driving frequency $\omega\gtrsim W$, where all
Floquet bands are clearly separated, or the case of small driving amplitude $t_F\ll \omega$, 
where the hierarchy of anticrossings can either be treated perturbatively in $t_F$ 
\cite{perez-piskunow_etal_pra_15} or one can use the 
rotating wave approximation (RWA) \cite{lindner_refael_galitski_natphys_11,lindner_etal_prb_13,
thakurathi_loss_klinovaja_prb_17,klinovaja_stano_loss_prl_16}.
In RWA one tunes $\omega$ in resonance with the gap $\Delta_g$ between the conduction and valence band 
such that only the conduction band and the first replica of 
the valence band are relevant. The perturbative regime of very small frequencies $\omega\ll\Delta_g$
has been treated in 
Refs.~\onlinecite{gomez_platero_prl_13,rodriguez-vega_seradjeh_prl_18,rodriguez-vega_etal_njp_18}.

In this Letter we consider solid state applications of FTIs in the generic 
regime, where all energy scales $t_F,\omega,\Delta_g$ are of comparable order and much smaller 
than $W$. In this regime, all Floquet replica-bands are strongly coupled and perturbative approaches
are not applicable. We will concentrate on the case of 1D systems with
chiral symmetry of the BDI class, however, our results have more general implications 
\cite{generic_case,SM}.  
As a testbed, we study the driven $2$-band Rashba nanowire in a Zeeman field $\Delta_Z$ proposed in 
Ref.~\onlinecite{thakurathi_loss_klinovaja_prb_17}, which, e.g., can be realized in curved bilayer graphene, 
where the band gap and the Rashba spin-orbit interaction (SOI) are tunable 
\cite{klinovaja_ferreira_loss_prb_2012}. For the BDI class, the topological
invariant is a winding number where an arbitrary positive number of TESs can be realized. By 
tuning $t_F$ and $\Delta_Z$, we will show that FTIs offer the possibility to access many of these 
topological phases. Most importantly, we find many regions in the phase diagram where 
several coexisting TESs appear of which some are strongly localized with localization length comparable
to the one appearing in the RWA phase. These TESs are stable and expected to be observable in experiments.
As a consequence, compared to static TIs and FTIs in the RWA regime, we find a dramatic extension of 
possible topological phase diagram regions. This strongly motivates experimental tests of the 
physics of FTIs at large driving amplitude in solid state systems and implies novel opportunities to 
detect unambiguous fingerprints of topological states.

To efficiently deal with the intriguing problem of strongly coupled Floquet bands,  
we propose an analysis based on the truncated Floquet Hamiltonian in frequency space. The strongly
localized TES of interest appear already in low truncation order and convergence is reached rapidly
even for large driving amplitude. Furthermore, in the presence of chiral symmetry, we show that
the number of boundary conditions can be reduced, in such a way that the number of TESs can be related 
straightforwardly to the number of roots of the determinant of the off-diagonal blocks of the 
Floquet Hamiltonian in the chiral basis as a function of complex quasimomentum. 
This allows for an elegant formulation of the bulk-boundary correspondence
together with the determination of the localization length of all TESs. 

{\it Model---} As a testbed we will use a tight-binding version of a recently proposed $2$-band 
Rashba nanowire where TESs appear when tuning a transverse Zeeman splitting $\Delta_Z$ 
or a driving field $t_F$, coupling the conduction and valence band \cite{thakurathi_loss_klinovaja_prb_17}.
The bulk Hamiltonian in quasimomentum space is given by
$\bar{h}_k(t) = (E_k + {\Delta_g\over 2} + \alpha_k \sigma_z)\eta_z + \Delta_Z\sigma_x + 2 t_F \cos(\omega t)\eta_x$, 
with quasimomentum $-\pi < k \le \pi$, dispersion relation $E_k=W\sin^2(k/2)$, and Rashba SOI
$\alpha_k=\alpha\sin{k}$.
Here, $\sigma_i$ and $\eta_i$ are Pauli matrices describing the spin and band degrees of freedom,
respectively. At resonance $\Delta_g=\omega$, we transform to the RWA basis by using the unitary transformation
$U(t)={1\over 2}\sum_{\pm}(1\pm\eta_z) e^{\mp i\omega t/2}$. The transformed Hamiltonian is given by 
\begin{align}
\label{eq:rwa_basis}
h_k(t) = h_k^R + t_F (\eta_+ e^{i\Omega t} + \eta_- e^{-i\Omega t}),
\end{align}
with $\eta_\pm={1\over 2}(\eta_x\pm i\eta_y)$ and an effective driving frequency $\Omega=2\omega$. Here,
$h_k^R=(E_k + \alpha_k\sigma_z)\eta_z + \Delta_Z\sigma_x + t_F\eta_x$ is the bulk Hamiltonian in RWA 
\cite{thakurathi_loss_klinovaja_prb_17}. It leads to a TES at zero energy for $t_F < \Delta_Z$ 
similar to the Majorana model proposed in Ref.~\onlinecite{oreg_refael_oppen_prl_10}.
However, beyond RWA, the second term on the r.h.s.~of Eq.~(\ref{eq:rwa_basis}) is an important 
correction term which can change the phase diagram significantly for larger values of $t_F,\Delta_Z\sim\Omega$. 

{\it Truncated Floquet Hamiltonian in frequency space---} 
The model defined by Eq.~(\ref{eq:rwa_basis}) falls into the class of generic time-periodic
1D systems with local chiral symmetry $S h_k(t) S = - h_k(-t)$, where $S=\sigma_z \eta_y$ for
our specific case. The Floquet Hamiltonian is defined by $(h^F_k)_{ll'} = h_{k,l-l'} - l\Omega\delta_{ll'}$, 
where $l=0,\pm 1,\pm 2,\dots$ denotes the Floquet modes and 
$h_{k,l}={1\over T}\int_0^T dt \,e^{il\Omega t}h_k(t)$ with $T={2\pi\over\Omega}$. 
It has a chiral symmetry $S_o h^F_k S_o = -h^F_k$ w.r.t.~zero energy, with $S_o=SP$ and 
$P|l\rangle=|-l\rangle$. 
Since the Floquet spectrum is defined $\text{modulo}\,\Omega$, there is also a 
chiral symmetry $S_e(h_k^F+{\Omega\over 2}) S_e = -(h_k^F+{\Omega\over 2})$ 
w.r.t.~energy $-{\Omega\over 2}$, with $S_e=ST_1P$ and
$T_1|l\rangle=|l+1\rangle$.
{ At $E=-{\Omega\over 2},0$ ($\text{modulo}\,\Omega$), two gaps open up due to the
couplings $\Delta_Z$ and $t_F$.} 
We are interested in TESs with
a localization length up to a certain threshold at these two energies. To analyze their occurrence,
we truncate $h^F_k$ by an even (odd) number $2 l_{\text{max}}$ ($2 l_{\text{max}}+1$) of Floquet replicas, 
where $l=-l_{\text{max}}+1, \dots,l_{\text{max}}$ ($l=-l_{\text{max}}, \dots,l_{\text{max}}$) are the allowed values
for the Floquet modes. This defines two different classes $h^{e/o,l_{\text{max}}}_k$ of finite-dimensional Floquet
Hamiltonians with chiral symmetries $S_{e/o}$, respectively. In real space, we obtain two corresponding 
tight-binding Floquet Hamiltonians $h^{e/o,l_{\text{max}}}_{nn'}$, where the integers 
$n,n'$ label the unit cells of dimension $2d$ (including spin, band, and Floquet modes).
The study of $h^{e/o,l_{\text{max}}}_{nn'}$ turns out to be very useful since the TESs of 
interest with shortest localization length already appear for small $l_{\text{max}}$ and convergence in
$l_{\text{max}}$ is reached rapidly even for large $t_F\sim\Omega$.

{\it Calculation of TESs---} 
We now construct TESs via an exact numerical approach for a half-infinite gapped system, 
with unit cells labeled by $n=1,2,\dots$, described by the Hamiltonians 
$h_{nn'}\equiv{\Omega\over 2}+h^{e,l_{\text{max}}}_{nn'}$ or $h_{nn'}\equiv h^{o,l_{\text{max}}}_{nn'}$,
in the even and odd case, respectively. The corresponding bulk Hamiltonian in $k$-space for the 
infinite system is denoted by $h_k$, which has a chiral symmetry $S\equiv S_{e,o}$ w.r.t.~zero energy. 
To find zero-energy TESs, which decay exponentially 
for $n\rightarrow\infty$, we determine the zero energy bulk eigenstates of $h_k$
with $\text{Im}(k)>0$, such that their linear combination vanishes at unit cell $n=0$ 
for all $2d$ bands. This procedure is applicable for all nearest-neighbor
hopping models like the one given by Eq.~(\ref{eq:rwa_basis}) \cite{nearest_neighbor}.
Since $D_k=\text{det}(h_k)$ is a power series in $z=e^{-ik}$ containing 
$z^{2d}, \dots, z^{-2d}$, there are in total $4d$ roots of $D_k=0$ in the complex plane 
of which $2d$ roots satisfy $\text{Im}(k)>0$ since $D_{k^*}=D_k^*$ 
[which follows from $h_{k^*}=(h_k)^\dagger$]. Therefore, the number of allowed zero 
energy bulk solutions is the same as the number of boundary conditions and it is not 
a priori clear whether the solutions are linearly dependent for $n=0$ to be able to fulfil the 
vanishing boundary condition. However, 
by exploiting chiral symmetry, we can write $h_k$ in the chiral basis by using the 
projectors $P_\pm={1\over 2}(1\pm S)$ on the states with chirality $\pm$. Since $P_\pm h_k P_\pm=0$, we find that 
$h_k$ is nondiagonal in the chiral basis with the $d\times d$-matrices 
$A^\pm_k=P_\mp h_k P_\pm$ defining the nondiagonal blocks. The TESs with chirality $\pm$ 
follow then from the above scheme by finding the roots of $D^\pm_k=\text{det}A^\pm_k=0$ with
$\text{Im}(k)>0$. This reduces the number of boundary conditions to $d$ and, since $D^\pm_k$
contains the powers $z^{d}, \dots, z^{-d}$, the total number of roots is given by $2d$. 
Crucially, the $2d$ roots can be distributed in an arbitrary manner between the upper and lower half of the complex
plane since $h_{k^*}=(h_k)^\dagger$ requires only $A^+_{k^*}=(A^-_k)^\dagger$ and $D^+_{k^*}=(D^-_k)^*$.  
Thus, if $N_\pm$ is the number of roots of $D^\pm_k=0$ with $\text{Im}(k)>0$, the boundary condition 
can be satisfied by at least $Z_\pm=\text{max}\{N_\pm-d,0\}$ possibilities determining the number of 
TESs with chirality $\pm$ (up to accidental cases not protected by chiral symmetry). We note that $N_\pm$ 
is a topological invariant since this number can only change when at least one of the roots crosses the real 
axis which corresponds to a closing of the bulk gap. Using $D^+_{k^*}=(D^-_k)^*$ we find that 
$N_\pm$ is the number of roots of $D^+_k=0$ with $\text{Im}(k)\gtrless 0$.
This gives $N_++N_-=2d$ or $N_+-d = -(N_--d)$ and we conclude that {\it all} TESs must have the same chirality 
and the total number of TESs is given by 
\begin{align} 
\label{eq:Z_edge}
Z = {1\over 2} |N_+ - N_-|.
\end{align}
The details of our construction of TESs for chiral 1D systems is provided in the
Supplemental Material (SM) containing also cases for special unit cells where the number of boundary
conditions can be further reduced \cite{SM}.
\begin{figure}[t]
\centering
 \includegraphics[width=\columnwidth]{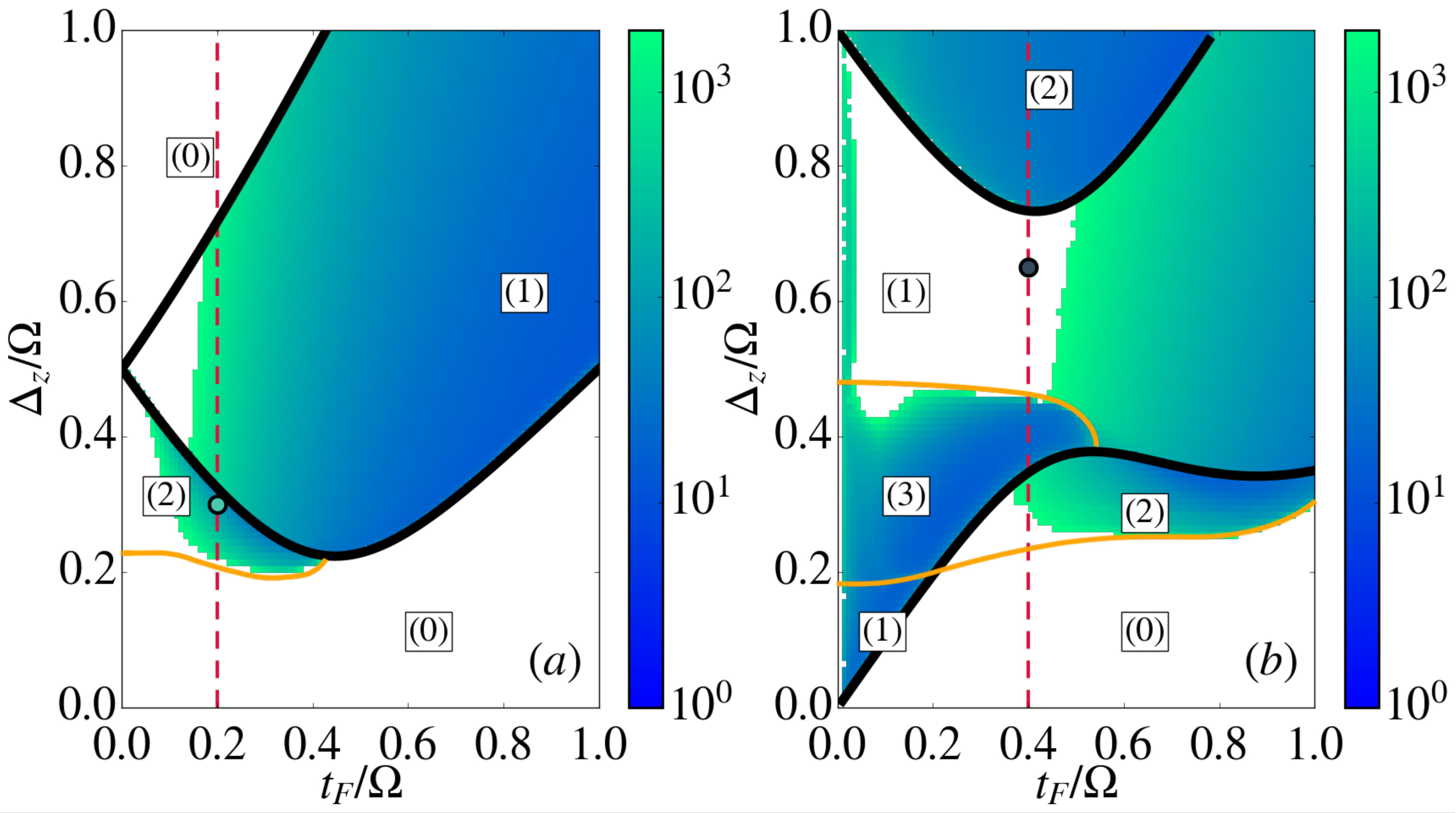}
 \caption{Phase diagrams for (a) 2 and (b) 3 Floquet replicas.
   The black (orange) phase boundaries indicate transitions where the number of TESs is changing by an 
   odd (even) number (so-called type ``A'' (``B'') phase transitions). 
   In each phase region the number of TESs is indicated in the boxes. The color indicates 
   the localization length $\xi$ of the strongest-localized TES. The regions are colored in white if no 
   TESs are present or if $\xi > 2000$. The vertical dashed lines indicate the
   values of $t_F$ along which the position of the roots of $D_k^+=0$ are shown in Fig.~\ref{fig:roots_2RWA}
   for $2$ Floquet replicas (for $3$ Floquet replicas see \cite{SM}).
   The dots indicate the parameter used in Fig.~\ref{fig:band_structure}.}
\label{fig:tes_23}
\end{figure}

In the SM, we also show that 
{ ${1\over 2}(N_+-N_-)$}
is identical to the winding number 
of $\text{det} A_k^+$ \cite{SM}, proving consistency
with other approaches \cite{schnyder_etal_njp_10}. Furthermore, by applying our scheme to 
$h_k\equiv{\Omega\over 2}+h^{e,l_{\text{max}}}_k$ and $h_k\equiv h^{o,l_{\text{max}}}_k$, we can 
calculate the number $Z^{e/o}_{l_{\text{max}}}$ of TESs at energy $E=-{\Omega\over 2}$ and $E=0$ 
via Eq.~(\ref{eq:Z_edge}). In the SM \cite{SM}, we show without any truncation that 
$Z^{e/o}=\lim_{l_{\text{max}}{\rightarrow\infty}}Z^{e/o}_{l_{\text{max}}}$ are identical to the 
{ absolute values of the}
topological invariants $\nu_{\pi}$/$\nu_0$
defined in Ref.~\onlinecite{asboth_tarasinski_delplace_prb_14} via special time 
evolution operators. While proving consistent with earlier works, our approach via a truncated Floquet
Hamiltonian in frequency space is essential to efficiently address the {\it{most important}} TESs, 
which are the ones with shortest localization length. Furthermore, we note that our scheme works as well for  
parabolic dispersion relations $E_k=k^2/(2m)$ or for linear Rashba terms $\alpha_k=\alpha k$ as 
they appear for continuum systems. 

\begin{figure}[!t]
\centering
 \includegraphics[width=\columnwidth]{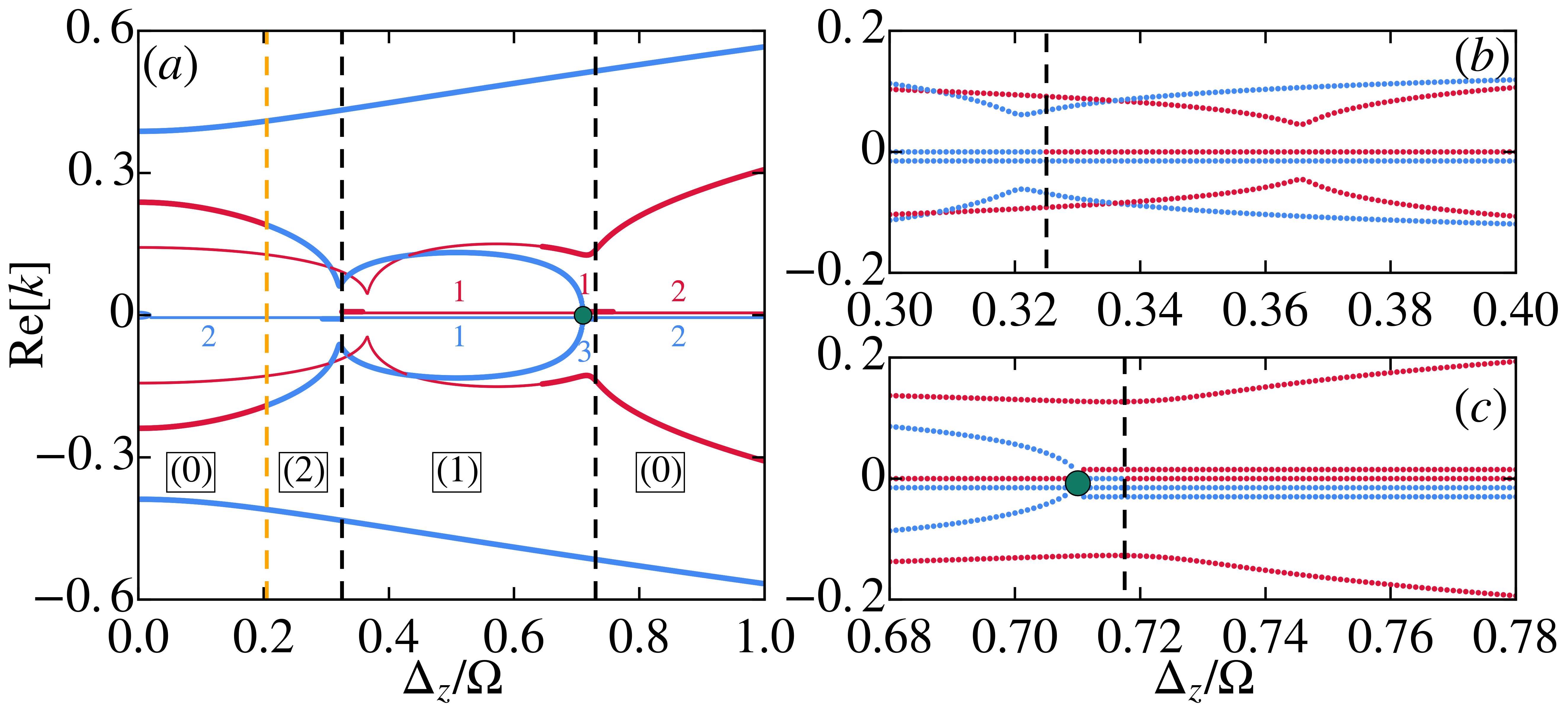}
 \caption{Real parts $\text{Re}(k_i)$ 
   of the roots of $D_{k_i}^+=0$ for $2$ Floquet replicas as a function of $\Delta_Z$ at
   fixed ${t_F\over\Omega}=0.2$ (i.e., along the dashed line in Fig.~\ref{fig:tes_23}(a)) . 
   Blue/red lines correspond to roots with $\text{Im}(k_i)\gtrless 0$.
   The vertical dashed lines correspond to phase transitions of type ``A'' (black) or ``B'' (orange). 
   The two transitions of type ``A'' are shown on a smaller scale in (b) and (c).
   If the real part of several roots is zero we indicate their number in (a) and use a slight offset
   in (b) and (c). Thick lines in (a) indicate roots with $|\text{Im}(k_i)|<0.14$,   which dominate TESs (if present) and can be identified 
   with the local minima in the bulk band structure (see Fig.~\ref{fig:band_structure}(a) 
   for ${\Delta_Z\over\Omega}=0.3$).  
   The boxes in (a) contain the number of TESs in the various phases. 
   The green dot in (a) and (c) indicates a bifurcation point which is very close to the phase transition.}
 \label{fig:roots_2RWA}
\end{figure}

\begin{figure}[t]
\centering
 \includegraphics[width=\columnwidth]{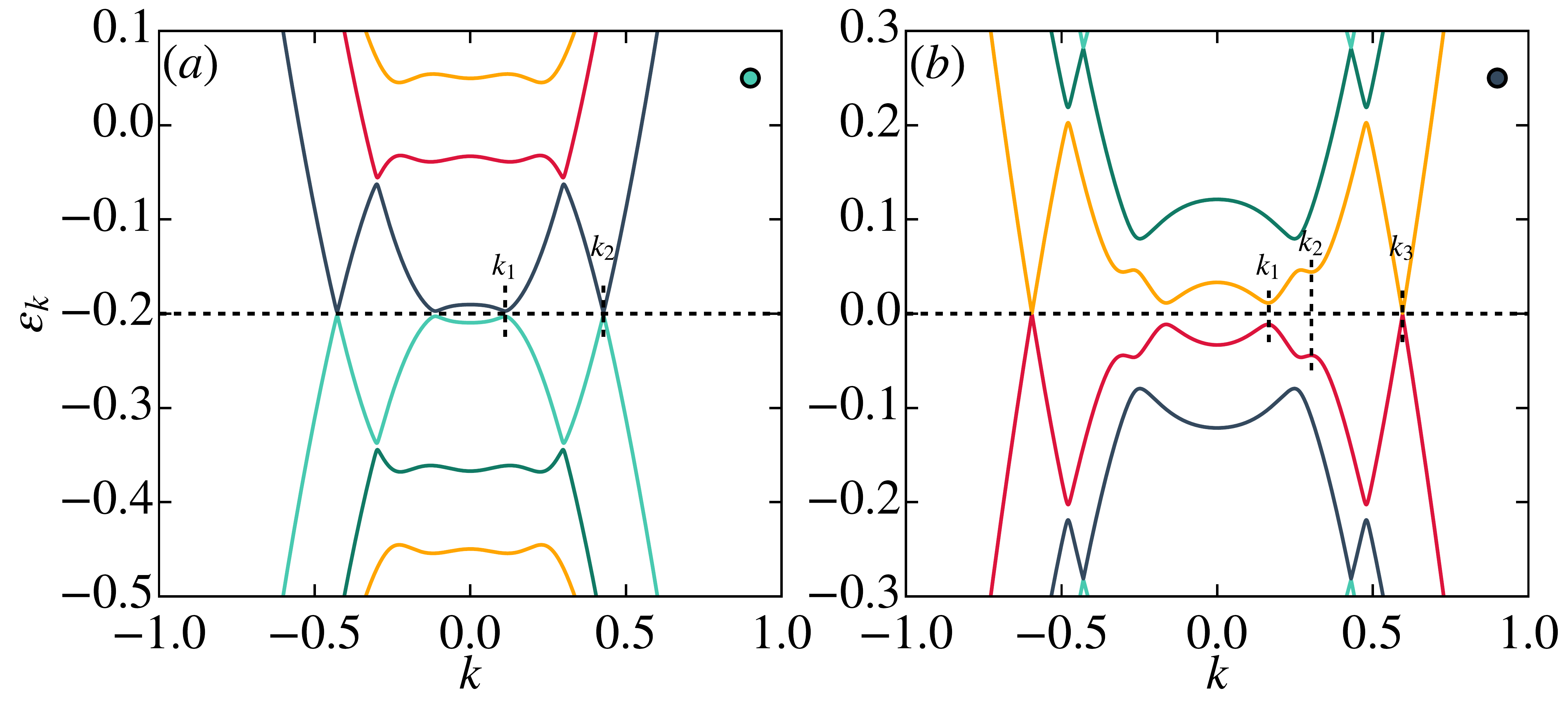}
 \caption{Band structure for (a) $2$ and (b) $3$ Floquet replicas 
   with (a) $({t_F\over\Omega},{\Delta_Z\over\Omega})=(0.2,0.3)$ and
   (b) $({t_F\over\Omega},{\Delta_Z\over\Omega})=(0.4,0.65)$ (corresponding to the parameter values
   at the dots in Fig.~\ref{fig:tes_23}). The label $k_i$ denote the positions of the local band minima
   which can be related to the real parts of the roots with small imaginary part in Fig.~\ref{fig:roots_2RWA}
   (and a corresponding figure in the SM for $3$ Floquet replicas \cite{SM}).
   The gaps at (a) $\pm k_{1,2}$ and (b) $\pm k_3$ are too small to be visible on the chosen scale.}
 \label{fig:band_structure}
\end{figure}

{\it Phase diagram---} Using the numerical approach described above we determine the 
TESs and the phase diagram as a function of $t_F$ and $\Delta_Z$. 
{ Corresponding to the two different gaps at $E=-{\Omega\over 2},0$ we show the phase
diagrams for an even/odd truncation of the Floquet replicas, respectively, and consider different 
truncation orders to analyse convergence.} 
For all figures we use the 
parameters $W=8$, $\Omega=0.4$, and $\alpha=0.3$. 
First, we use a low (not yet converged) truncation order $l_{\text{max}}=1$.   
The number of TESs and the localization length \cite{loc} of the strongest-localized
TES is shown in Fig.~\ref{fig:tes_23}. To understand the phase boundaries, 
we show in Fig.~\ref{fig:roots_2RWA} the evolution of the real part $\text{Re}(k)$ of the roots of $D_k^+=0$ 
along the vertical dashed line in Fig.~\ref{fig:tes_23}(a) (an analog figure is shown in the SM along the dashed 
line in Fig.~\ref{fig:tes_23}(b) \cite{SM}), together with the sign of $\text{Im}(k)$
(indicated by blue/red color for $\pm$). There are two different classes of phase boundaries in Fig.~\ref{fig:tes_23} 
(in the following called ``A'' and ``B''), 
where the number $Z$ of TESs is either changing by an odd (black lines) or even (orange lines) number. 
A transition of type ``A'' occurs for a root crossing $k=0$ along the imaginary axis which changes the
number $Z$ of TESs by one cf.~Eq.~(\ref{eq:Z_edge}). A transition of type ``B'' 
is associated with two roots crossing simultaneously the real axis at finite $\pm k_i$ 
\cite{time_reversal_symmetry} leading to a change of $Z$ by two. 
At the phase transition, the bulk gap closes either at $k=0$ (type ``A'') or at finite $\pm k_i$ (type ``B'').
Therefore, phase boundaries of type ``A'' are independent of the SOI, whereas the ones of type ``B'' depend 
on $\alpha$. Moreover, phase transitions of type ``A'' are often but not always associated with a bifurcation
point lying quite close to the phase boundary (indicated by a green dot in
Fig.~\ref{fig:roots_2RWA}(a,c)). A bifurcation point occurs when two roots with real 
parts $\pm k_i\ne 0$ merge on the imaginary axis and, subsequently, move along the
imaginary axis in different directions such that one of the roots crosses through $k=0$ at the phase
transition. We note that, sufficiently apart from the phase 
boundaries and for sufficiently small values of $t_F$ and $\Delta_Z$, the real parts of the roots with small 
imaginary part (thick lines in Fig.~\ref{fig:roots_2RWA}) are related to the positions $k_i$ 
of the local minima of the bulk band structure, see Fig.~\ref{fig:band_structure}(a). Furthermore, 
the energy distance $\Delta_i$ of the bulk energy $\epsilon_{k_i}$
from the gap center agrees roughly (up to a factor of $O(1)$) with the imaginary part $\kappa_i$ 
of the roots via $\kappa_i\sim{\Delta_i\over v_i}$, with $v_i=(\partial_k E)_{k_i}={W\over 2}\sin(k_i)$.
However, for larger values of $t_F$ and $\Delta_Z$ as indicated in Fig.~\ref{fig:band_structure}(b), the  
Floquet bands are strongly coupled leading to a significant 
{ broadening}
of the anticrossings and minima 
at small $k$, and there is no possibility to set up effective theories treating each anticrossing separately and
coupling them perturbatively.    

Of particular interest are the phase diagram regions where
the strongest-localized TES has a small localization length (blue regions in Fig.~\ref{fig:tes_23})
and is formed from roots with small real part (where the imaginary part is larger, see also 
the band structure shown in Fig.~\ref{fig:band_structure}(b)). 
In this case the strongest-localized TES can coexist with other TESs of larger localization length 
since they have all the same chirality 
{ (also called \lq\lq sublattice'')}
and can not repel each other. This important 
feature persists for higher truncation order. 

\begin{figure}
\centering
 \includegraphics[width=\columnwidth]{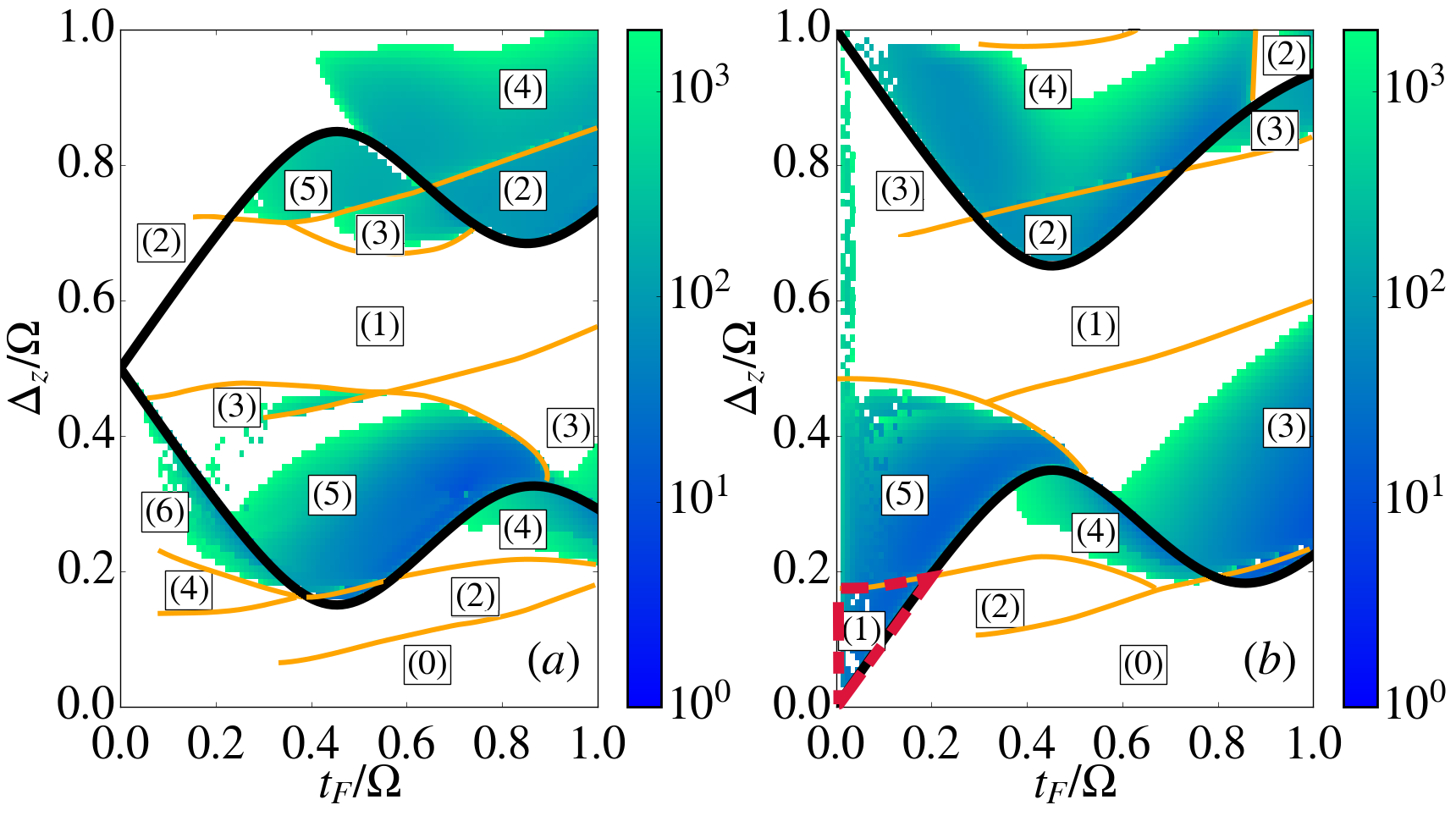}
 \caption{Phase diagrams for (a) 6 and (b) 5 Floquet replicas (same notations as in Fig.~\ref{fig:tes_23}).
   All phase boundaries shown are almost converged and do not change significantly by
   increasing the truncation order (some of them end at small $t_F$ due to numerical instabilities). 
   The red-dashed contour in (b) indicates the RWA phase where $t_F,\Delta_Z\ll\Omega$.}
 \label{fig:tes_65}
\end{figure}

The phase diagram for a truncation with $6$ and $5$ Floquet replicas is shown in Fig.~\ref{fig:tes_65}.
We have presented only those phase boundaries which do not change significantly by increasing the truncation order.
This does not exclude that further pairs of TESs can arise from very small gaps at large $k$ not captured
in the considered truncation order leading to additional phase boundaries. However, these TESs have a very
large localization length far beyond those ones occurring in Fig.~\ref{fig:tes_65} and are of minor interest.
Therefore, the most important feature in the phase diagram is {\it not} 
the total number of TESs (which converges slowly) but the identification of those regions where the 
strongest-localized TES has a small localization length (which converges always very fast). 
Besides the RWA phase [indicated by a red-dashed contour
in Fig.~\ref{fig:tes_65}(b)], there are many new regions in the phase diagram with TESs of comparable 
localization length to the one occurring in the RWA phase. Furthermore, we find that the RWA phase 
is extended to the surprisingly large region
$t_F<\Delta_Z\lesssim {\Omega\over 2}$ but for $0.2\lesssim\Delta_z\lesssim{\Omega\over 2}$ additional
TESs are present besides the strongly localized RWA state. 
The strongly localized TESs occurring already at a low truncation order are stable, 
since a closing of very small gaps at anticrossings with large $k$ will give the TESs only a very
small broadening but will not influence their localization length significantly. 

We have complemented our results by an exact diagonalization study on a finite system with $N=10000$ unit cells showing consistency as well as the occurrence of finite energy nontopological edge states


%

{\it Conclusions---} 
We have analyzed the topological properties of chiral 1D FTIs beyond RWA and have developed 
a method to deal with the occurrence of many anticrossings of Floquet replicas.
This unique feature of FTIs is shown to lead to many new regions in the topological
phase diagram with TESs of very short localization length accessible to experiments. 
Regions with multiple TESs with the same chirality could be useful as realizations of qudits 
(which are d-dimensional extensions of qubits). We analyzed driven Rasha nanowires as a concrete 
Floquet model, which we propose can be put to an experimental test in curved bilayer graphene, 
where both the models parameters, band gap and the Rashba term, can be tuned by external parameters 
\cite{klinovaja_ferreira_loss_prb_2012}. 
{ The presence of TESs can be detected most accurately via STM transport spectroscopy 
which probes resonances of the spectral density inside the gap and is not sensitive to the rather 
subtle occupation of the Floquet states. Moreover, our proposed approach via a truncated Floquet
Hamiltonian may also be useful for a further development of other methods analysing topological
states like, e.g., the scattering state formalism \cite{fulga_maksymenko_prb_16}.}


\begin{acknowledgments}
We thank V. Meden and S. Wessel for fruitful discussions. 
This work was supported by the Deutsche Forschungsgemeinschaft via RTG 1995, the Swiss National 
Science Foundation (SNSF) and NCCR QSIT. Simulations were performed with computing resources 
granted by RWTH Aachen University under projects rwth0347, rwth0362, thes0445, and prep0010.
Funding was received from the European Union's Horizon 2020 research, innovation program (ERC
Starting Grant, grant agreement No 757725) as well as from the independence grant from the TR 183 network.
\end{acknowledgments}


\begin{thebibliography}{99}

\bibitem{kitagawa_etal_10}
T. Kitagawa, E. Berg, M. Rudner, and E. Demler, Phys. Rev. B {\bf 82}, 235114 (2010).

\bibitem{kitagawa_etal_pra_10}
T. Kitagawa, M.S. Rudner, E. Berg, and E. Demler, Phys. Rev. A {\bf 82}, 033429 (2010).

\bibitem{rudner_lindner_berg_levin_prx_13}
M.S. Rudner, N.H. Lindner, E. Berg, and M. Levin, Phys. Rev. X {\bf 3}, 031005 (2013).

\bibitem{asboth_obuse_prb_13}
J.K. Asb\'oth and H. Obuse, Phys. Rev. B {\bf 88}, 121406(R) (2013).

\bibitem{asboth_tarasinski_delplace_prb_14}
J.K. Asb\'oth, B. Tarasinski, and P. Delplace, Phys. Rev. B {\bf 90}, 125143 (2014).

\bibitem{nathan_rudner_njp_15}
F. Nathan and M. S. Rudner, New J. Phys. {\bf 17}, 125014 (2015).

\bibitem{carpentier_delplace_fruchart_gawedzki_prl_15}
D. Carpentier, P. Delplace, M. Fruchart, and K. Gawedzki, Phys. Rev. Lett. {\bf 114}, 106806 (2015).

\bibitem{fulga_maksymenko_prb_16}
I.C. Fulga and M. Maksymenko, Phys. Rev. B {\bf 93}, 075405 (2016).

\bibitem{fruchart_prb_16}
M. Fruchart, Phys. Rev. B {\bf 93}, 115429 (2016).

\bibitem{roy_harper_prb_17}
R. Roy and F. Harper, Phys. Rev. B {\bf 95}, 195128 (2017); 
{\it ibid.} {\bf 96}, 155118 (2017). 

\bibitem{yao_etal_prb_17}
S. Yao, Z. Yan and Z. Wang, Phys. Rev. B {\bf 96}, 195303 (2017).

\bibitem{hoeckendorf_etal_prb_18}
B. H\"ockendorf, A. Alvermann, and H. Fehske, Phys. Rev. B {\bf 97}, 045140 (2018).




\bibitem{kitagawa_etal_natcomm_12}
T. Kitagawa, M.A. Broome, A. Fedrizzi, M.S. Rudner, E. Berg, I. Kassal, 
A. Aspuru-Guzik, E. Demler, and A.G. White, Nat. Commun. 3, {\bf 882} (2012).

\bibitem{rechtsman_etal_nature_13}
M.C. Rechtsman, J.M. Zeuner, Y. Plotnik, Y. Lumer, D. Podolsky, F. Dreisow, 
S. Nolte, M. Segev, and A. Szameit, Nature (London) {\bf 496}, 196 (2013).

\bibitem{lu_etal_natphoton_14}
L. Lu, J.D. Joannopoulos, and M. Soljacic, Nat. Photon. {\bf 8}, 821 (2014).

\bibitem{cardano_etal_natcomm_17}
F. Cardano, A. D'Errico, A. Dauphin, M. Maffei, B. Piccirillo, C. de Lisio, G. De Filippis, 
V. Cataudella, E. Santamato, L. Marrucci, M. Lewenstein, and P. Massignan, 
Nat. Commun. {\bf 8}, 15516 (2017).

\bibitem{maczewsky_etal_natcomm_17}
L.J. Maczewsky, J.M. Zeuner, S. Nolte, and A. Szameit, 
Nat. Commun. {\bf 8}, 13756 (2017).

\bibitem{mukherjee_etal_natcomm_17}
S. Mukherjee, A. Spracklen, M. Valiente, E. Andersson, P. \"Ohberg, N. Goldman, and R.R. Thomson,
Nat. Comm. {\bf 8}, 13918 (2017).




\bibitem{jotzu_etal_nature_14}
G. Jotzu, M. Messer, R. Desbuquois, M. Lebrat, T. Uehlinger, D. Greif, and T. Esslinger, 
Nature (London) {\bf 515}, 237 (2014).

\bibitem{jimenez-garcia_etal_prl_15}
K. Jim\'enez-Garcia, L.J. LeBlanc, R.A. Williams, M.C. Beeler, C. Qu, M. Gong, C. Zhang, and I.B. Spielman, 
Phys. Rev. Lett. {\bf 114}, 125301 (2015).

\bibitem{quelle_etal_njp_17}
A. Quelle, C. Weitenberg, K. Sengstock, and C. Morais Smith,
New J. Phys. {\bf 19}, 113010 (2017).




\bibitem{calvo_etal_apl_11}
H.L. Calvo, H.M. Pastawski, S. Roche, and L.E.F. Foa Torres, 
Appl. Phys. Lett. {\bf 98}, 232103 (2011). 

\bibitem{suarez_torres_prb_12}
E. Su\'arez Morell and L.E.F. Foa Torres, Phys. Rev. B {\bf 86}, 125449 (2012).

\bibitem{wang_etal_science_13}
Y.H. Wang, H. Steinberg, P. Jarillo-Herrero, and N. Gedik, Science {\bf 342}, 453 (2013).

\bibitem{usaj_etal_prb_14}
G. Usaj, P.M. Perez-Piskunow, L.E.F. Foa Torres, and C.A. Balseiro,
Phys. Rev. B {\bf 90}, 115423 (2014).

\bibitem{sie_etal_natmat_15}
E.J. Sie, J.W. McIver, Yi-Hsien Lee, L. Fu, J. Kong, and N. Gedik, 
Nature Materials {\bf 14}, 290 (2015).

\bibitem{calvo_etal_prb_15}
H.L. Calvo, L.E.F. Foa Torres, P.M. Perez-Piskunow, C.A. Balseiro, and G. Usaj,
Phys. Rev. B {\bf 91}, 241404 (2015).

\bibitem{wang_scirep_17}
Y. Wang, Y. Liu, and B. Wang, Sci. Rep. {\bf 7}, 41644 (2017).



\bibitem{jiang_etal_prl_11}
L. Jiang, T. Kitagawa, J. Alicea, A.R. Akhmerov, D. Pekker, G. Refael, J.I. Cirac, E. Demler, M.D. Lukin,
and P. Zoller, Phys. Rev. Lett. {\bf 106}, 220402 (2011).

\bibitem{kundu_seradjeh_prl_13}
A. Kundu and B. Seradjeh, Phys. Rev. Lett. {\bf 111}, 136402 (2013).

\bibitem{reynoso_frustaglia_prb_13}
A.A. Reynoso and D. Frustaglia, Phys. Rev. B {\bf 87}, 115420 (2013).

\bibitem{thakurathi_etal_prb_13}
M. Thakurathi, A.A. Patel, D. Sen, and A. Dutta, Phys. Rev. B {\bf 88}, 155133 (2013).

\bibitem{thakurathi_sengupta_sen_prb_14}
M. Thakurathi, K. Sengupta, and D. Sen, Phys. Rev. B {\bf 89}, 235434 (2014).


\bibitem{thakurathi_loss_klinovaja_prb_17}
M. Thakurathi, D. Loss, and J. Klinovaja, Phys. Rev. B {\bf 95}, 155407 (2017).


\bibitem{oka_aoki_prb_09}
T. Oka and H. Aoki, Phys. Rev. B {\bf 79}, 081406 (2009).

\bibitem{inoue_tanaka_prl_10}
J.-I. Inoue and A. Tanaka, Phys. Rev. Lett. {\bf 105}, 017401 (2010).

\bibitem{lindner_refael_galitski_natphys_11}
N.H. Lindner, G. Refael, and V. Galitski, Nat. Phys. {\bf 7}, 490 (2011).

\bibitem{dora_etal_prl_12}
B. D\'ora, J. Cayssol, F. Simon, and R. Moessner, Phys. Rev. Lett. {\bf 108}, 056602 (2012).

\bibitem{perez-piskunow_etal_pra_15}
P.M. Perez-Piskunow, L.E.F. Foa Torres, and G. Usaj,
Phys. Rev. A {\bf 91}, 043625 (2015); in this reference also the case $t_F,\omega\gtrsim 0.1 W$ has been studied which is orthogonal to our parameter regime.


\bibitem{lindner_etal_prb_13}
N.H. Lindner, D.L. Bergman, G. Refael, and V. Galitski, Phys. Rev. B {\bf 87}, 235131 (2013).


\bibitem{klinovaja_stano_loss_prl_16}
J. Klinovaja, P. Stano, and D. Loss, Phys. Rev. Lett. {\bf 116}, 176401 (2016).



\bibitem{hbar}
We set $\hbar$ and the lattice spacing to unity throughout this work. 



\bibitem{gomez_platero_prl_13}
A. G\'{o}mez-Le\'{o}n and G. Platero, Phys. Rev. Lett. {\bf 110}, 200403 (2013).

\bibitem{rodriguez-vega_seradjeh_prl_18}
M. Rodriguez-Vega and B. Seradjeh, Phys. Rev. Lett. {\bf 121}, 036402 (2018).

\bibitem{rodriguez-vega_etal_njp_18}
M. Rodriguez-Vega, M. Lentz, and B. Seradjeh, New J. Phys. {\bf 20}, 093022 (2018).



\bibitem{generic_case}
Our method can be applied to the class $CII$ and $AIII$ as well and further generalizations will be described in a future work, see also \cite{SM}.

\bibitem{SM}
See Supplemental Material. 

\bibitem{klinovaja_ferreira_loss_prb_2012}
J. Klinovaja, G.J. Ferreira, and D. Loss, Phys. Rev. B {\bf 86}, 235416 (2012).

\bibitem{oreg_refael_oppen_prl_10}
Y. Oreg, G. Refael, and F. von Oppen, Phys. Rev. Lett. {\bf 105}, 177002 (2010).

\bibitem{nearest_neighbor}
This can always be achieved by taking the unit cell large enough. 

\bibitem{loc}
The localization length $\xi$ of the TESs
is calculated from the inverse participation ratio (IPR). 

\bibitem{schnyder_etal_njp_10}
S. Ryu, A.P. Schnyder, A. Furusaki, and A.W.W. Ludwig,
New J. Phys. {\bf 12}, 065010 (2010).

\bibitem{time_reversal_symmetry}
Due to time-reversal symmetry of the model Eq.~(\ref{eq:rwa_basis}), 
the roots with finite real part appear in pairs at $k_i$ and $-k_i^*$.  

\end{thebibliography}
\end{document}


\title{Chiral 1D Floquet topological insulators beyond rotating wave approximation\\
Supplemental Material}

\author{Dante M. Kennes}
\affiliation{Dahlem Center for Complex Quantum Systems and Fachbereich Physik, 
Freie Universit\"at Berlin, 14195 Berlin, Germany}
\author{Niclas M\"uller}
\affiliation{Institut f\"ur Theorie der Statistischen Physik, RWTH Aachen, 
52056 Aachen, Germany and JARA - Fundamentals of Future Information Technology}
\author{Mikhail Pletyukhov}
\affiliation{Institut f\"ur Theorie der Statistischen Physik, RWTH Aachen, 
52056 Aachen, Germany and JARA - Fundamentals of Future Information Technology}
\author{Clara Weber}
\affiliation{Institut f\"ur Theorie der Statistischen Physik, RWTH Aachen, 
52056 Aachen, Germany and JARA - Fundamentals of Future Information Technology}
\author{Christoph Bruder}
\affiliation{Department of Physics, University of Basel, Klingelbergstrasse 82, CH-4056 Basel, Switzerland}
\author{Fabian Hassler}
\affiliation{JARA-Institute Quantum Information, RWTH Aachen University, 52056 Aachen, Germany}
\author{Jelena Klinovaja}
\affiliation{Department of Physics, University of Basel, Klingelbergstrasse 82, CH-4056 Basel, Switzerland}
\author{Daniel Loss}
\affiliation{Department of Physics, University of Basel, Klingelbergstrasse 82, CH-4056 Basel, Switzerland}
\author{Herbert Schoeller}
\email[Email: ]{schoeller@physik.rwth-aachen.de}
\affiliation{Institut f\"ur Theorie der Statistischen Physik, RWTH Aachen, 
52056 Aachen, Germany and JARA - Fundamentals of Future Information Technology}

\maketitle

The following appendices comprise the Supplemental Material to
Ref.~\onlinecite{kennes_etal}: (1) We will present the proof of the bulk-boundary correspondence and the
construction of the topological edge states (TESs) in the general case together with the relation 
to the concept of topological invariants defined via winding numbers for the static as well as 
for the dynamic (Floquet) case. (2) We relate the phase diagram for
$3$ Floquet replicas to the evolution of the roots of the determinant of the
nondiagonal blocks of the Hamiltonian in the chiral basis. (3) We present the results from the
exact diagonalization study for a finite system and discuss the occurrence of nontopological 
edge states.

\section{Bulk-boundary correspondence and construction of TESs}
\label{sec:bulk_boundary_tes}

A generic tight-binding model for a half-infinite chain with an even number of bands (required for 
chiral symmetry) can be written as
\begin{align}
\label{eq:tb_half}
H\,=\,\sum_{n=1}^\infty \left\{|n\rangle\langle n| \otimes h(0) \,+\,
|n+1\rangle\langle n| \otimes h(1) \,+\,|n\rangle\langle n+1| \otimes h(-1) \right\}\,,
\end{align}
where $n=1,2,\dots$ labels the unit cells with $2d$ bands and $h(\delta)=h(-\delta)^\dagger$, $\delta=0,\pm 1$, 
are $2d\times2d$-matrices describing
the hopping processes from unit cell $n$ to $n+\delta$. We have taken the unit cell large enough such that only 
nearest-neighbor hopping processes occur. Assuming a local chiral symmetry $S=S^\dagger=S^{-1}$ such that
\begin{align}
\label{eq:S}
S h(\delta) S \,=\, -h(\delta) \quad,\,\text{for}\quad \delta=0,\pm 1 \,,
\end{align}
we can decompose $h(\delta)$ as
\begin{align}
\label{eq:A_delta}
h(\delta)\,=\,A^+(\delta) \,+\,A^-(\delta) \quad,\quad 
A^+(\delta)\,=\,P_- h(\delta) P_+ \quad,\quad
A^-(\delta)\,=\,P_+ h(\delta) P_- \,=\, A^+(-\delta)^\dagger \,,
\end{align}
where $P_\pm={1\over 2}(1\pm S)$ project on the subspaces with chirality $\pm$. In the same
way we can split the total Hamiltonian of the half-infinite chain as
\begin{align}
\label{eq:A}
H\,=\,A^+ \,+\,A^- \quad,\quad 
A^+\,=\,P_- H P_+ \quad,\quad
A^-\,=\,P_+ H P_- \,=\, (A^+)^\dagger \,,
\end{align}
with
\begin{align}
\label{eq:A_half}
A^\pm\,=\,\sum_{n=1}^\infty \left\{|n\rangle\langle n| \otimes A^\pm(0) \,+\,
|n+1\rangle\langle n| \otimes A^\pm(1) \,+\,|n\rangle\langle n+1| \otimes A^\pm(-1) \right\}\,.
\end{align}
In addition we define the corresponding quantities for the bulk where we sum over all unit cells
\begin{align}
\label{eq:H_bulk}
H_{\text{bulk}}\,&=\,\sum_{n=-\infty}^\infty \left\{|n\rangle\langle n| \otimes h(0) \,+\,
|n+1\rangle\langle n| \otimes h(1) \,+\,|n\rangle\langle n+1| \otimes h(-1) \right\}\,,\\
\label{eq:A_bulk}
A^\pm_{\text{bulk}}\,&=\,\sum_{n=-\infty}^\infty \left\{|n\rangle\langle n| \otimes A^\pm(0) \,+\,
|n+1\rangle\langle n| \otimes A^\pm(1) \,+\,|n\rangle\langle n+1| \otimes A^\pm(-1) \right\}\,,
\end{align}
which can be written in quasimomentum space 
$|k\rangle={1\over\sqrt{2\pi}}\sum_{n=-\infty}^\infty e^{ikn}|n\rangle$, $-\pi < k \le \pi$, in the 
diagonal form
\begin{align}
\label{eq:HA_bulk_k}
H_{\text{bulk}}\,=\,\int_{-\pi}^\pi dk\, |k\rangle\langle k| \otimes h_k \quad,\quad
A^\pm_{\text{bulk}}\,=\,\int_{-\pi}^\pi dk\, |k\rangle\langle k| \otimes A^\pm_k \,,
\end{align}
with 
\begin{align}
\label{eq:HA_k}
h_k\,=\,\sum_{\delta=0,\pm 1} e^{-ik\delta} h(\delta) \,=\, (h_k)^\dagger \,=\, A^+_k\,+\,A^-_k \quad,\quad
A^\pm_k\,=\,\sum_{\delta=0,\pm 1} e^{-ik\delta} A^\pm(\delta) \,=\, (A^\mp_k)^\dagger\,.
\end{align}

The goal is to find a TES $|\psi^\pm\rangle$ with zero energy $H|\psi^\pm\rangle=0$
and chirality $S|\psi^\pm\rangle=\pm|\psi^\pm\rangle$ exponentially decaying 
for $n\rightarrow\infty$. Using (\ref{eq:A}) this means that we have to solve the equations
\begin{align}
\label{eq:tes_equation}
A^+ |\psi^+\rangle\,=\,0 \quad,\quad A^- |\psi^-\rangle\,=\,0 \,.
\end{align}
To construct the TES we make the following ansatz in unit cell $n$
\begin{align}
\label{eq:tes_ansatz}
|\psi^\pm(n)\rangle \,=\, \sum_{i=1}^{N_\pm} \,\lambda_i^\pm\,|\psi_{k_i}^\pm(n)\rangle \,,
\end{align}
where $|\psi_{k_i}^\pm\rangle$ are solutions of the bulk problem with complex quasimomentum $k_i$ in
the upper half of the complex plane $\text{Im}k_i>0$, and zero energy
\begin{align}
\label{eq:tes_form}
|\psi_{k_i}^\pm(n)\rangle \,=\, |\chi_{k_i}^\pm\rangle \, e^{i k_i n} \quad,\quad
A^\pm_{k_i}|\chi_{k_i}^\pm\rangle\,=\,0 \,.
\end{align}
Inserting the ansatz (\ref{eq:tes_ansatz}) in (\ref{eq:tes_equation}) we get with (\ref{eq:A_half}) for
unit cell $n\ge 1$
\begin{align}
\nonumber
\langle n|A^\pm |\psi^\pm\rangle \,
&=\, \sum_{m=1}^\infty \sum_{\delta=0,\pm 1}
\langle n|m+\delta\rangle A^\pm(\delta) |\psi^\pm(m)\rangle \\
\nonumber
&=\, \delta_{n1} \sum_{\delta=0,-1} A^\pm(\delta) |\psi^\pm(1-\delta)\rangle 
\,+\,\delta_{n>1} \sum_{\delta=0,\pm} A^\pm(\delta) |\psi^\pm(n-\delta)\rangle \\
\nonumber
&=\, -\delta_{n1} A^\pm(1) |\psi^\pm(0)\rangle 
\,+\,\sum_{\delta=0,\pm} A^\pm(\delta) |\psi^\pm(n-\delta)\rangle \\
\nonumber
&=\, -\delta_{n1} A^\pm(1) |\psi^\pm(0)\rangle 
\,+\,\sum_{i=1}^{N_\pm}\lambda^\pm_i e^{i k_i n} 
\sum_{\delta=0,\pm} A^\pm(\delta) e^{-ik_i\delta}|\chi^\pm_{k_i}\rangle \\
\nonumber
&=\, -\delta_{n1} A^\pm(1) |\psi^\pm(0)\rangle 
\,+\,\sum_{i=1}^{N_\pm}\lambda^\pm_i e^{i k_i n} A^\pm_{k_i}|\chi^\pm_{k_i}\rangle \\
\label{eq:insert_ansatz}
&=\, -\delta_{n1} A^\pm(1) |\psi^\pm(0)\rangle \,,
\end{align}
where we have used (\ref{eq:tes_ansatz}), (\ref{eq:tes_form}), and (\ref{eq:HA_k}) in the last three steps. 
Thus, we have to fulfil the following boundary condition on an artificial unit cell at $n=0$
\begin{align}
\label{eq:boundary_condition}
A^\pm(1) |\psi^\pm(0)\rangle \,=\, 
A^\pm(1) \sum_{i=1}^{N_\pm} \lambda^\pm_i |\chi^\pm_{k_i}\rangle \,=\, 0 \,.
\end{align}
This boundary condition is obviously fulfilled for $\sum_{i=1}^{N_\pm} \lambda^\pm_i |\chi^\pm_{k_i}\rangle =0$ but
the number of boundary conditions can be reduced if the $d\times d$-matrices $A^+(1)$ and 
$A^-(1)=A^+(-1)^\dagger$ contain zero matrix elements, which are connected to some symmetries assumed
for the model under consideration. If by permuting rows and columns, one can find $r_\pm$ rows
and $s_\pm$ columns of $A^+(\pm 1)$ which are zero (possibly after some common unitary transformation
which commutes with the chiral symmetry), then the number of boundary conditions is given
by $d_\pm=\text{min}\{r_\pm,s_\pm\}$. These numbers are directly related to the powers of $\text{det}A^+_k$ in
$z=e^{-ik}$, since $A^+_k=A^+(0)+e^{-ik}A^+(1)+e^{ik}A^+(-1)$, which gives
\begin{align}
\label{eq:powers_detA}
\text{det} A^+_k\,\rightarrow\, z^{d_+} \,,\,z^{d_+-1} \,,\,\dots\,,\, z^{-d_-+1} \,,\,z^{-d_-} \,.
\end{align}
The total number of solutions of (\ref{eq:tes_form}) is given by the number $N_\pm$ of solutions of 
$\text{det}A^\pm_k=0$ for $k$ in the upper half of the complex plane, i.e. $\text{Im}(k)>0$ and
$-\pi < \text{Re}(k) \le \pi$. Since $d_\pm$ boundary conditions have to be fulfilled this means
that it is guaranteed that at least
\begin{align}
\label{eq:Z_pm}
Z_\pm \,=\, \text{max}\{N_\pm-d_\pm,0\} 
\end{align}
solutions can be found for the set of coefficients $\{\lambda^\pm_i\}_i$ in (\ref{eq:tes_ansatz}), up
to additional accidental solutions which are not protected by symmetry. $Z_\pm$ is the number of 
TESs with chirality $\pm$. Since $A^-_k=(A^+_{k^*})^\dagger$ for complex $k$ (see Eq.~(\ref{eq:HA_k}) 
for real $k$) we get $\text{det}A^-_k=(\text{det}A^+_{k^*})^*$, i.e. the number
of solutions of $\text{det}A^-_k=0$ in the upper half of the complex plane is identical to
the number of solutions of $\text{det}A^+_k=0$ in the lower half of the complex plane. We conclude 
that $N_\pm$ is the number of solutions of $\text{det}A^+_k=0$ for $\text{Im}(k)\gtrless 0$. Using 
(\ref{eq:powers_detA}), the total number $N_++N_-$ of solutions of $\text{det}A^+_k=0$ in the complex 
plane with $-\pi < \text{Re}(k) \le \pi$ is given by $d_++d_-$, which gives the important equation
\begin{align}
\label{eq:N_d}
N_+ - d_+ \,=\, - (N_- - d_-) \,.
\end{align} 
Together with (\ref{eq:Z_pm}) we find the final result for the total number $Z$ of TESs
\begin{align}
\label{eq:Z}
Z \,=\, Z_+ + Z_- \,=\, {1\over 2}|N_+-N_--(d_+-d_-)| \,.
\end{align}

We note that the construction of the TESs via (\ref{eq:tes_ansatz}) and (\ref{eq:tes_form}) does
not work if several solutions of $\text{det}A^+_k=0$ coincide (i.e. the case of multiple roots). 
At such special bifurcation points the ansatz for the TES has to be an exponential together with
a pre-exponential polynomial. However, this are very special points in the phase diagram with 
mass zero and the formula (\ref{eq:Z}) for the number $Z$ of TESs is not changed if one includes in
$N_\pm$ the multiplicity of each root. 

Our approach can as well be used for the chiral classes $CII$ and $AIII$ in $1D$. In class $CII$ Kramers
pairs will occur for all TESs due to time-reversal symmetry with $T^2=-1$. In class $AIII$ neither 
time-reversal nor particle-hole symmetry is fulfilled. This has the only consequence that the solutions
of Eq.~(\ref{eq:tes_form}) do no longer occur in time-reversed pairs $k_i$ and $-k_i^*$, with the effect
that a topological phase transition changing the parity of the number of TESs can also happen at finite $k$. 

In case chiral symmetry is not fulfilled or if the number of roots of $\text{det}A_k^+=0$ with 
$\text{Im}(k_i)\gtrless 0$ is the same (as can be shown for the chiral classes $DIII$ and $CI$), it is also 
possible to study the TESs by calculating all $N=2d$ solutions of 
\begin{align}
\label{eq:chi_k_generic}
h_{k_i}|\chi_{k_i}\rangle\,=\,0\quad,\quad \text{Im}(k_i)\,>\,0\,,
\end{align}
and constructing the wave function of the TES by 
\begin{align}
\label{eq:TES_generic}
|\psi(n)\rangle\,=\,\sum_{i=1}^N\,\lambda_i\,|\psi_{k_i}\rangle\quad,\quad
|\psi_{k_i}\rangle\,=\,|\chi_{k_i}\rangle\,e^{i k_i n}\,,
\end{align}
provided that the boundary condition $\sum_{i=1}^N\lambda_i|\chi_{k_i}\rangle=0$ can be fulfilled. 
Although the number $N$ of solutions is identical to the dimension of the vectors $\chi_{k_i}$, this
can be easily checked numerically by analysing the rank $r$ of the matrix of all eigenvectors $\chi_{k_i}$.
If $r<2d$ the eigenvectors are linearly dependent and the boundary condition can be fulfilled. 

Furthermore, we note that also the topological classes $D$ and $DIII$ in $1D$ with $\mathbbm{Z}_2$-symmetry can
be studied by a variant of our proposed method for chiral systems and higher dimensions can be analysed by
dimensional reduction. This goes beyond the scope of the present work and requires a systematic study and
classification of all topological systems which will be part of future works \cite{future_work}.

Our proposed method to calculate the TESs via the bulk properties of the system relies essentially on translational invariance along the periodic lattice. In the presence of disorder translational invariance is broken and our method can not be applied. However, the stability of a certain TES can still be studied by comparing the disorder-induced broadening $\Gamma$ with the corresponding gap $\Delta$ of the dispersion relation in which the TES is sitting. If $\Gamma\ll\Delta$, the TES is coupled only weakly to the bulk spectrum and its stability will be preserved. For weak disorder, a rough estimation of $\Gamma$ can be obtained via golden rule in perturbation theory. This requires also the knowledge of all bulk states in the absence of disorder. These states can also be calculated by our method. A bulk solution is again of the form (\ref{eq:TES_generic}) but in addition there are two solutions of (\ref{eq:chi_k_generic}) with $\text{Im}(k_i)=0$, besides $2d-1$ solutions with $\text{Im}(k_i)>0$ and $2d-1$ solutions with $\text{Im}(k_i)<0$ (giving $4d$ solutions in total). This means that there are in total $2d+1$ solutions with $\text{Im}(k_i)\ge 0$ which is sufficient to fulfil $2d$ boundary conditions.

\section{Winding numbers for chiral static Hamiltonians}
\label{sec:winding_static}

The number $Z$ of TESs given by Eq.~(\ref{eq:Z}) via the number $N_\pm$ of roots of the 
function $f_k=\text{det}A^+_k$ in the upper/lower half of the complex plane 
for $-\pi < \text{Re}(k) \le \pi$ can be related
to its winding number $\nu$ around the origin when $k$ changes on the real axis from $-\pi\rightarrow \pi$
\begin{align}
\label{eq:winding}
\nu\,=\,{1\over 2\pi i}\int_{-\pi}^{\pi} dk {f^\prime_k\over f_k} \,,
\end{align}
where $f^\prime_k={df_k\over dk}$ is the derivative and the winding is defined positive for 
counterclockwise rotation here. In the following we will show that
\begin{align}
\label{eq:winding_Z}
\nu\,=\, N_+ - d_+ \,=\, -(N_--d_-) \quad \Rightarrow \quad |\nu| \,=\, Z \,,
\end{align}
where we have used (\ref{eq:N_d}) and (\ref{eq:Z}) for the second and third identity. Therefore it
is sufficient to prove that $\nu=N_+ - d_+$.

To find this relation we use the argument principle which gives the following relation for the
number of roots of an analytic function $f_k$ inside any closed contour $\gamma$ in the complex plane
\begin{align}
\label{eq:argument_principle}
N_\gamma \,=\, {1\over 2\pi i}\oint_\gamma dk {f^\prime_k\over f_k} \,,
\end{align}
where the direction of $\gamma$ is counterclockwise. Choosing for $\gamma$ a closed curve consisting
of four straight lines according to 
$-\pi\rightarrow \pi \rightarrow \pi+i\infty \rightarrow -\pi+i\infty \rightarrow -\pi$, we surround 
obviously all roots of $f_k$ in the upper half, i.e. $N_\gamma=N_+$. Using (\ref{eq:winding}) the
path from $-\pi\rightarrow\pi$ gives the winding number $\nu$. Since $f_k$ is periodic in 
$k\rightarrow k+2\pi$, the two contributions from $\pi\rightarrow\pi+i\infty$ and 
$-\pi+i\infty\rightarrow -\pi$ cancel each other. Thus, we get
\begin{align}
\label{eq:Np}
N_+ \,=\, \nu \,-\,{1\over 2\pi i} \int_{-\pi+i\infty}^{\pi+i\infty} dk {f^\prime_k\over f_k}\,.
\end{align}
For $\text{Im}(k)\rightarrow\infty$ we find from (\ref{eq:powers_detA}) that $f_k=a e^{-ikd_+}$ with
some $k$-independent prefactor $a$. This leads to ${f^\prime_k\over f_k}=-id_+$ and the second term 
on the r.h.s. of (\ref{eq:Np}) gives $d_+$. This gives the desired relation $\nu=N_+-d_+$.  

The winding number can be expressed in many alternative ways. Using 
${f^\prime_k\over f_k}={d\over dk}\ln(\text{det}A^+_k)={d\over dk}\text{Tr}\ln(A^+_k)$ we can write
\begin{align}
\label{eq:winding_trace}
\nu\,=\,{1\over 2\pi i} \int_{-\pi}^\pi \text{Tr} (A^+_k)^{-1} d A^+_k \,.
\end{align}
Instead of $A^+_k$ one can also use the unitary $q$-matrix $q_k={1\over\sqrt{A^+_k(A^+_k)^\dagger}}A^+_k$.
Since the winding number of a hermitian matrix is zero we get
\begin{align}
\label{eq:winding_trace_q}
\nu\,=\,{1\over 2\pi i} \int_{-\pi}^\pi \text{Tr} \, q_k^\dagger d q_k \,.
\end{align}
The $q$-matrix can be expressed in a compact way via the representation of the Hamiltonian in a
basis of eigenstates $|x^\alpha_{k\sigma}\rangle$ ($\alpha=1,\dots,d\,$; $\sigma=\pm$) such that 
\begin{align} 
\label{eq:h_basis}
h_k |x^\alpha_{k\sigma}\rangle \,=\, \sigma \epsilon^\alpha_k |x^\alpha_{k\sigma}\rangle \quad,\quad
S |x^\alpha_{k\sigma}\rangle \,=\, |x^\alpha_{k\bar{\sigma}}\rangle \,,
\end{align}
with $\bar{\sigma}=-\sigma$ and $\epsilon^\alpha_k > 0$. 
This is always possible for a gapped Hamiltonian with chiral symmetry. In this basis the projectors
$P_\pm = {1\over 2} (1\pm S)$ act as follows
\begin{align}
\label{eq:P_properties}
P_+ |x^\alpha_{k\sigma}\rangle \,=\, P_+ |x^\alpha_{k+}\rangle \quad,\quad
P_- |x^\alpha_{k\sigma}\rangle \,=\, \sigma P_- |x^\alpha_{k+}\rangle \,.
\end{align}
This gives for the matrix $A^+_k = P_- h_k P_+$ the result
\begin{align}
\label{eq:A_basis}
A^+_k \,&=\,\sum_{\alpha\sigma} \sigma\epsilon^\alpha_k P_- |x^\alpha_{k\sigma}\rangle\langle x^\alpha_{k\sigma}| P_+ \\
&=\,2 \sum_{\alpha} \epsilon^\alpha_k P_- |x^\alpha_{k+}\rangle\langle x^\alpha_{k+}| P_+ \,,
\end{align}
and the following result for the $q$-matrix
\begin{align}
\label{eq:q_basis}
q_k \,=\,2 \sum_{\alpha} P_- |x^\alpha_{k+}\rangle\langle x^\alpha_{k+}| P_+ \,.
\end{align}
Using $2 \langle x^\alpha_{k+}|P_\sigma |x^{\alpha'}_{k+}\rangle = \delta_{\alpha \alpha'}$ we find after
some straightforward algebra the following useful result for the integrand of the winding number 
(\ref{eq:winding_trace_q})
\begin{align}
\label{eq:winding_q_basis}
\text{Tr} \,q_k^\dagger d q_k \,=\, 2 \sum_\alpha \langle x^\alpha_{k+}|S| d x^\alpha_{k+}\rangle 
\,=\, 2 \sum_\alpha \langle x^\alpha_{k-}|S| d x^\alpha_{k-}\rangle \,,
\end{align}
where we used $S^2=1$ and $S|x^\alpha_{k+}\rangle=|x^\alpha_{k-}\rangle$ in the last step.

\section{Winding numbers for chiral Floquet Hamiltonians}
\label{sec:winding_floquet}

We start from a time-dependent periodic Hamiltonian $h_k(t)=h_k(t+T)=h_k(t)^\dagger$ with oscillation frequency
$\Omega={2\pi \over T}$ in quasimomentum representation which has the local chiral symmetry
\begin{align}
\label{eq:S_time}
S h_k(t) S \,=\, - h_k(-t) \,.
\end{align}
The corresponding Floquet-Hamiltonian $h^F_k$ is defined by
\begin{align}
\label{eq:floquet_h}
(h^F_k)_{ll'} \,=\, h_{k,l-l'}\,-\,l\Omega\delta_{ll'} \,,
\end{align}
where 
\begin{align}
\label{eq:floquet_modes}
|l\rangle\,=\,\int_0^T dt |t\rangle\langle t|l\rangle \quad,\quad 
\langle t|l\rangle\,=\,{1\over\sqrt{T}} \,e^{-il\Omega t}
\end{align}
denote the Floquet modes and 
\begin{align}
\label{eq:h_l}
h_{k,l}\,=\,{1\over T}\,\int_0^T dt \,e^{il\Omega t}\,h_k(t) \,=\, h_{k,-l}^\dagger
\end{align}
are the discrete Fourier components of $h_k(t)$ which fulfil the chiral symmetry
\begin{align}
\label{eq:h_l_symmetries}
S h_{k,l} S \,=\, - h_{k,-l} \,.
\end{align}

Due to the Bloch theorem all solutions of the time-dependent Schr\"odinger equation of $h_k(t)$ can be 
determined from the eigenvectors $|x_k\rangle$ of the Floquet Hamiltonian $h^F_k$ with
quasienergy $\epsilon_k$ as 
\begin{align}
\label{eq:solution_SG}
\psi_k(t)\,=\,\sqrt{T}\,e^{-i\epsilon_k t} |x_k(t)\rangle \,,
\end{align}
where 
\begin{align}
\label{eq:x_t}
|x_k(t)\rangle \,=\, \langle t|x_k\rangle \,=\, \sum_l\,\langle t|l\rangle\langle l|x_k\rangle
\,=\, {1\over\sqrt{T}}\,\sum_l\,e^{-il\Omega t} |x_{k,l}\rangle 
\end{align}
is the Fourier transform of $|x_{k,l}\rangle=\langle l|x_k\rangle$ w.r.t. to the Floquet modes. The
quasienergy is defined uniquely up to multiples of $\Omega$ since the Floquet Hamiltonian transforms in the
following way under translations $T_r|l\rangle=|l+r\rangle$ of the Floquet modes
\begin{align}
\label{eq:translational_trafo}
T_r h_k^F T_{-r} \,=\, h_k^F\,+\, r\Omega \,.
\end{align}
This means that 
\begin{align}
\label{eq:shift_omega}
h^F_k\,|x_k\rangle\,=\,\epsilon_k\,|x_k\rangle \quad \Rightarrow \quad
h^F_k\,T_{-r}|x_k\rangle\,=\,(\epsilon_k + r\Omega)\,T_{-r}|x_k\rangle \,.
\end{align}

Due to (\ref{eq:floquet_h}), (\ref{eq:h_l_symmetries}) and (\ref{eq:translational_trafo}) 
we find that $h_k^F$ has two 
independent chiral symmetries $S_F$ and $\bar{S}_F$ w.r.t. the two energies $E=0$ and $E=-{\Omega\over 2}$ 
(both $\text{modulo}\,\Omega$) defined by
\begin{align}
\label{eq:def_chiral_symmetries}
S_F\,=\,PS \quad,\quad \bar{S}_F\,=\,T_1 P S \,,
\end{align}
where $P|l\rangle = |-l\rangle$ changes the parity of the Floquet modes. 
This means that with $\bar{h}^F_k=h^F_k+{\Omega\over 2}$ we get 
\begin{align}
\label{eq:h_property_chiral}
S_F h^F_k S_F \,=\, - h^F_k \quad,\quad \bar{S}_F \bar{h}^F_k \bar{S}^F_k \,=\, - \bar{h}^F_k \,.
\end{align}
We note that the chiral symmetry remains even if the Floquet Hamiltonian $h^F_k$ ($\bar{h}^F_k$)
is truncated with an odd (even) number of Floquet modes $l=-l_{\text{max}},\dots,l_{\text{max}}$
($l=-l_{\text{max}}+1,\dots,l_{\text{max}}$).  
According to (\ref{eq:winding_trace_q}) this allows the definition of two winding numbers via
two $q$-matrices $q^F_k$ and $\bar{q}^F_k$ corresponding to $S_F$ and $\bar{S}_F$, respectively 
\begin{align}
\label{eq:winding_nu_floquet}
\nu\,=\,{1\over 2\pi i} \int_{-\pi}^\pi \text{Tr} \, (q^F_k)^\dagger d q^F_k \quad,\quad
\bar{\nu}\,=\,{1\over 2\pi i} \int_{-\pi}^\pi \text{Tr} \, (\bar{q}^F_k)^\dagger d \bar{q}^F_k \,.
\end{align}
Due to (\ref{eq:winding_Z}) these topological invariants correspond to the number of TESs
at energies $E=0$ and $E=-{\Omega\over 2}$ even for a truncated Floquet Hamiltonian. For finite 
truncation it is impossible to relate the invariants to other invariants defined in the literature 
via the time evolution operators $U_k(T,0)$ and $U_k({T\over 2},-{T\over 2})$ of $h_k(t)$, 
where no truncation is used \cite{asboth_tarasinski_delplace_prb_14}. Therefore, we consider 
in the following the limit $l_{\text{max}}\rightarrow\infty$ in order to see that $\nu$ and $\bar{\nu}$ 
can be written as
\begin{align}
\label{eq:relation_winding}
\nu\,=\,{1\over 2}(\nu^{\prime} + \nu^{\prime\prime}) \quad,\quad
\bar{\nu}\,=\,{1\over 2}(\nu^{\prime} - \nu^{\prime\prime}) \,,
\end{align}
where $\nu^{\prime}$ and $\nu^{\prime\prime}$ are two winding numbers defined w.r.t. two chiral Hamiltonians
$h^{\prime}_k$ and $h^{\prime\prime}_k$ which follow from the time evolution operators as
\begin{align}
\label{eq:h_h_prime_via_U}
U_k(T,0)\,=\, e^{-i h^{\prime}_k T} \quad,\quad 
U_k({T\over 2},-{T\over 2})\,=\, e^{-i h^{\prime\prime}_k T} \,.
\end{align}
Eq.~(\ref{eq:relation_winding}) is the definition of $\nu_0=\nu$ and $\nu_\pi=\bar{\nu}$ used in
Ref.~\onlinecite{asboth_tarasinski_delplace_prb_14}, where it was argued that these invariants define
the number of TESs at energy $E=0$ and $E=-{\Omega\over 2}$, respectively. 

To prove (\ref{eq:relation_winding}) for an untruncated Floquet Hamiltonian we first define 
a convenient basis $\{|x^{\alpha r}_{k\sigma}\rangle\}$ in Floquet space, analog to (\ref{eq:h_basis})
\begin{align} 
\label{eq:h_F_basis}
h^F_k |x^{\alpha r}_{k\sigma}\rangle \,=\, \sigma (\epsilon^\alpha_k + r\Omega) |x^{\alpha r}_{k\sigma}\rangle 
\quad &,\quad
S_F |x^{\alpha r}_{k\sigma}\rangle \,=\, |x^{\alpha r}_{k\bar{\sigma}}\rangle \,,\\
\label{eq:bar_h_F_basis}
\bar{h}^F_k |\bar{x}^{\alpha r}_{k\sigma}\rangle \,=\, 
\sigma (\bar{\epsilon}^\alpha_k + r\Omega) |\bar{x}^{\alpha r}_{k\sigma}\rangle 
\quad &,\quad
\bar{S}_F |\bar{x}^{\alpha r}_{k\sigma}\rangle \,=\, |\bar{x}^{\alpha r}_{k\bar{\sigma}}\rangle \,,
\end{align}
where $0<\epsilon^\alpha_k,\bar{\epsilon}^\alpha_k\le {\Omega\over 2}$, 
$\alpha=1,\dots,d$, $\sigma=\pm$, $r=0,\pm 1,\pm 2,\dots$, and 
\begin{align}
\label{eq:x_floquet}
|x^{\alpha r}_{k\sigma}\rangle\,=\,T_{\bar{\sigma}r}|x^{\alpha}_{k\sigma}\rangle \quad,\quad
|\bar{x}^{\alpha r}_{k\sigma}\rangle\,=\,T_{\bar{\sigma}r}|\bar{x}^{\alpha}_{k\sigma}\rangle \,,
\end{align}
together with $|x^\alpha_{k\sigma}\rangle=|x^{\alpha,r=0}_{k\sigma}\rangle$, 
$|\bar{x}^\alpha_{k\sigma}\rangle=|\bar{x}^{\alpha,r=0}_{k\sigma}\rangle$ and
\begin{align}
\label{eq:bar_x_r=0}
\bar{\epsilon}^\alpha_k\,=\,-\epsilon^\alpha_k\,+\,{\Omega\over 2} \quad,\quad
|\bar{x}^{\alpha}_{k+}\rangle\,=\,|x^\alpha_{k-}\rangle \quad,\quad
|\bar{x}^{\alpha}_{k-}\rangle\,=\,T_1|x^\alpha_{k+}\rangle \,.
\end{align}

Using (\ref{eq:winding_q_basis}), we can write for the integrands of the winding numbers in
Eq.~(\ref{eq:winding_nu_floquet})
\begin{align}
\label{eq:winding_q_basis_floquet}
\text{Tr} \,(q^F_k)^\dagger d q^F_k \,=\, 2 \sum_{\alpha r} \langle x^{\alpha r}_{k+}|S_F| d x^{\alpha r}_{k+}\rangle 
\quad,\quad
\text{Tr} \,(\bar{q}^F_k)^\dagger d \bar{q}^F_k \,=\, 
2 \sum_{\alpha r} \langle \bar{x}^{\alpha r}_{k-}|\bar{S}_F| d \bar{x}^{\alpha r}_{k-}\rangle
\,.
\end{align}
Inserting (\ref{eq:h_property_chiral}) and (\ref{eq:x_floquet}), and using (\ref{eq:bar_x_r=0}), we find
\begin{align}
\nonumber
\text{Tr} \,(q^F_k)^\dagger d q^F_k \,&=\, 
2 \sum_{\alpha r} \langle x^{\alpha}_{k+}|T_r P T_{-r} S| d x^{\alpha}_{k+}\rangle 
\,=\,2 \sum_{\alpha r} \sum_{ll'}\langle x^{\alpha}_{k+,l}|(T_r P T_{-r})_{ll'} S| d x^{\alpha}_{k+,l'}\rangle \\
\nonumber
&=\,2 \sum_{\alpha r} \sum_{ll'} \delta_{l-r,-l'+r} \langle x^{\alpha}_{k+,l}|S| d x^{\alpha}_{k+,l'}\rangle 
\,=\,2 \sum_{\alpha} \sum_{\substack{ll'\\ l+l' \,\text{even}}} \langle x^{\alpha}_{k+,l}|S| d x^{\alpha}_{k+,l'}\rangle \\
\label{eq:winding_odd}
&=\,\sum_{\alpha} \sum_{ll'} \left\{1 + (-1)^{l+l'}\right\}\langle x^{\alpha}_{k+,l}|S| d x^{\alpha}_{k+,l'}\rangle 
\,=\,\sum_{\alpha} \left\{\langle y^{\alpha}_{k+}|S| d y^{\alpha}_{k+}\rangle \,+\,
\langle \tilde{y}^{\alpha}_{k+}|S| d \tilde{y}^{\alpha}_{k+}\rangle \right\}\,,\\
\nonumber
\text{Tr} \,(\bar{q}^F_k)^\dagger d \bar{q}^F_k \,&=\, 
2 \sum_{\alpha r} \langle x^{\alpha}_{k+}|T_{-1}T_{-r} T_1 P T_r T_1 S| d x^{\alpha}_{k+}\rangle 
\,=\,2 \sum_{\alpha r} \sum_{ll'}\langle x^{\alpha}_{k+,l}|(T_{-r} P T_{r+1})_{ll'} S| d x^{\alpha}_{k+,l'}\rangle \\
\nonumber
&=\,2 \sum_{\alpha r} \sum_{ll'} \delta_{l+r,-l'-r-1} \langle x^{\alpha}_{k+,l}|S| d x^{\alpha}_{k+,l'}\rangle 
\,=\,2 \sum_{\alpha} \sum_{\substack{ll'\\ l+l' \,\text{odd}}} \langle x^{\alpha}_{k+,l}|S| d x^{\alpha}_{k+,l'}\rangle \\
\label{eq:winding_even}
&=\,\sum_{\alpha} \sum_{ll'} \left\{1 - (-1)^{l+l'}\right\}\langle x^{\alpha}_{k+,l}|S| d x^{\alpha}_{k+,l'}\rangle 
\,=\,\sum_{\alpha} \left\{\langle y^{\alpha}_{k+}|S| d y^{\alpha}_{k+}\rangle \,-\,
\langle \tilde{y}^{\alpha}_{k+}|S| d \tilde{y}^{\alpha}_{k+}\rangle \right\}\,,
\end{align}
where we have defined
\begin{align}
\label{eq:y_tilde_y}
|y^\alpha_{k\sigma}\rangle \,=\, \sum_l\,|x^\alpha_{k\sigma,l}\rangle 
\,=\, \sqrt{T}\,|x^\alpha_{k\sigma}(t=0)\rangle \quad,\quad 
|\tilde{y}^\alpha_{k\sigma}\rangle \,=\, \sum_l\,(-1)^l\,|x^\alpha_{k\sigma,l}\rangle 
\,=\, \sqrt{T}\,|x^\alpha_{k\sigma}(t={T\over 2})\rangle \,.
\end{align}
\begin{figure}[t]
\centering
 \includegraphics[height=5cm]{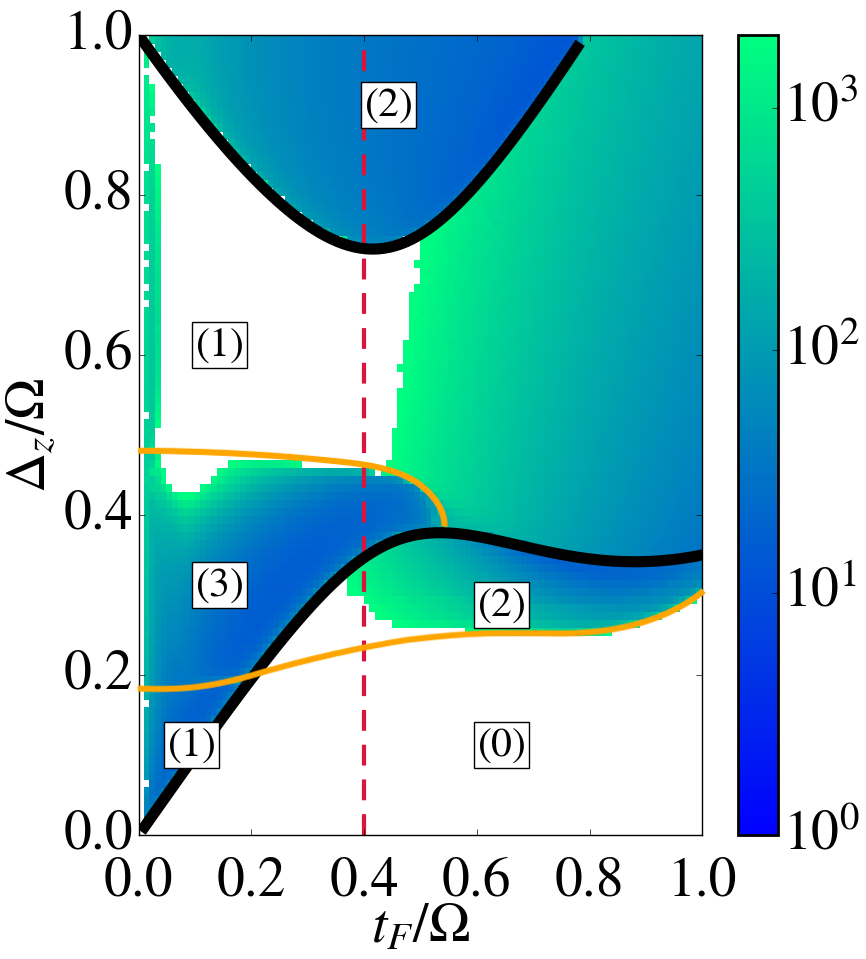}
 \caption{Phase diagram for 3 Floquet replicas.
   The black (orange) phase boundaries indicate transitions where the number of TESs is changing by an 
   odd (even) number (so-called type ``A'' (``B'') phase transitions). 
   In each phase region the number of TESs is indicated in the boxes. The color indicates 
   the localization length $\xi$ of the strongest-localized TES. The regions are colored in white if no 
   TESs are present or if $\xi > 2000$. The vertical dashed line indicates the
   value of $t_F$ along which the position of the roots of $D_k^+=\text{det} A_k^+=0$ 
   are shown in Fig.~\ref{fig:roots_3RWA}.}
\label{fig:tes_3}
\end{figure}
\begin{figure}[!t]
\centering
 \includegraphics[height=8cm]{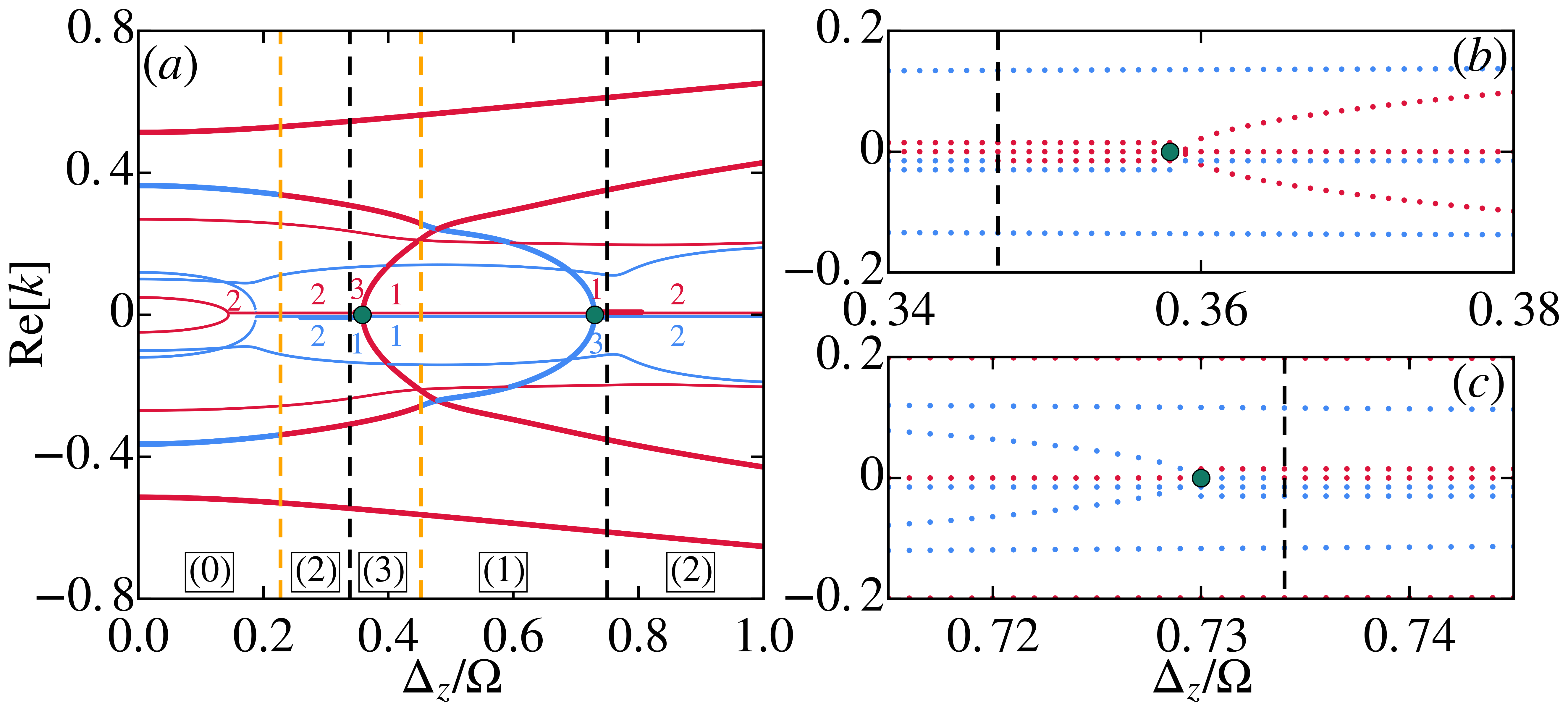}
 \caption{The evolution of the real parts $\text{Re}(k_i)$ 
   of the roots of $D_{k_i}^+=\text{det}A_{k_i}^+=0$ for $3$ Floquet replicas as a function of $\Delta_Z$ at
   fixed ${t_F\over\Omega}=0.4$ (i.e., along the dashed line in Fig.~\ref{fig:tes_3}) . 
   Blue/red lines correspond to roots with $\text{Im}(k_i)\gtrless 0$.
   The vertical dashed lines correspond to phase transitions of type ``A'' (black) or ``B'' (orange). 
   The two transitions of type ``A'' are shown on a smaller scale in (b) and (c).
   If the real part of several roots is zero we indicate their number in (a) and use a slight offset
   in (b) and (c). Thick lines in (a) indicate roots with $|\text{Im}(k_i)|<0.14$.
   The boxes in (a) contain the number of TESs in the various phases. 
   The green dots indicate bifurcation points which are very close to the phase transitions.}
 \label{fig:roots_3RWA}
\end{figure}
As a result we find with (\ref{eq:winding_q_basis}) that Eq.~(\ref{eq:relation_winding}) holds, 
where $\nu^\prime$ and $\nu^{\prime\prime}$ 
are the winding numbers of two Hamiltonians $h^\prime_k$ and $h^{\prime\prime}_k$, respectively, 
which have chiral symmetry $S$ and are formally defined by 
\begin{align}
\label{eq:h_h_prime}
h^\prime_k \,=\, \sum_{\alpha\sigma} \sigma\epsilon^\alpha_k \,
|y^\alpha_{k\sigma}\rangle\langle y^\alpha_{k\sigma}| \quad,\quad 
h^{\prime\prime}_k \,=\, \sum_{\alpha\sigma} \sigma\epsilon^\alpha_k \,
|\tilde{y}^\alpha_{k\sigma}\rangle\langle \tilde{y}^\alpha_{k\sigma}| \,.
\end{align}
These Hamiltonians are identical to the ones defined via (\ref{eq:h_h_prime_via_U}) since 
due to (\ref{eq:solution_SG}) the time evolution operator can be written as
\begin{align}
\label{eq:U_t_t_prime}
U_k(t,t')\,=\, T \sum_{\alpha\sigma} \,e^{-i\sigma\epsilon^\alpha_k(t-t')}\, 
|x^\alpha_{k\sigma}(t)\rangle\langle x^\alpha_{k\sigma}(t')| \,.
\end{align} 
\begin{figure}
\centering
 \includegraphics[height=5cm]{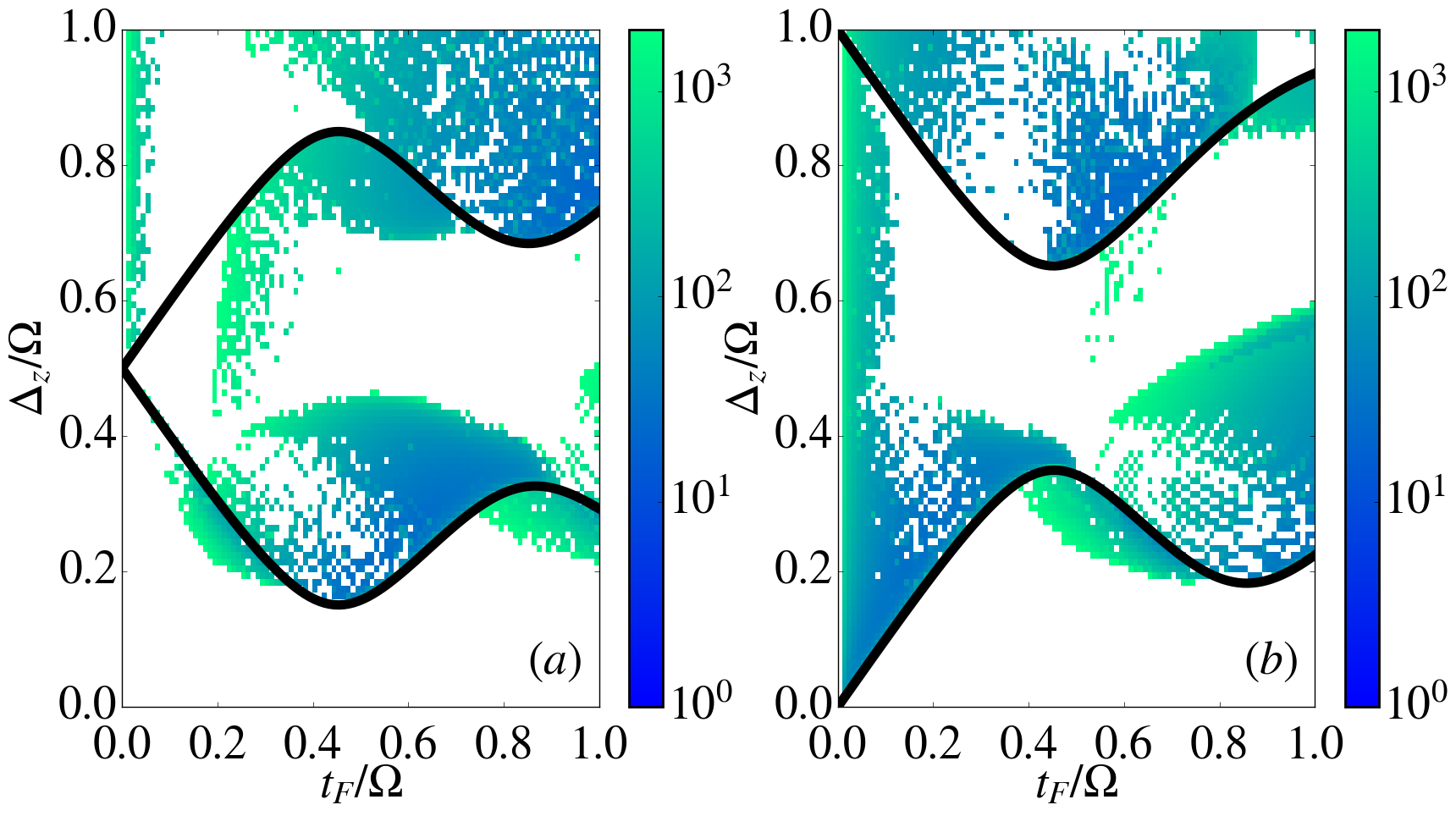}
 \caption{Exact diagonalization study for the tight-binding Floquet Hamiltonians 
   (a) $h^{e,3}_{nn'}$ (6 replicas) and (b) $h^{o,2}_{nn'}$ (5 replicas) on a finite system 
   with $N=10000$ unit cells. The parameters are $W=8$, $\alpha=0.3$ and $\Omega=0.4$. 
   The color indicates the localization length of the strongest-localized edge state. Only 
   those edge states are considered which lie deeply
   in the bulk gap (within $15\%$ away from the gap center relative to the bulk gap). }
\label{fig:exdia_tes_65}
\end{figure}
\begin{figure}
\centering
 \includegraphics[height=5cm]{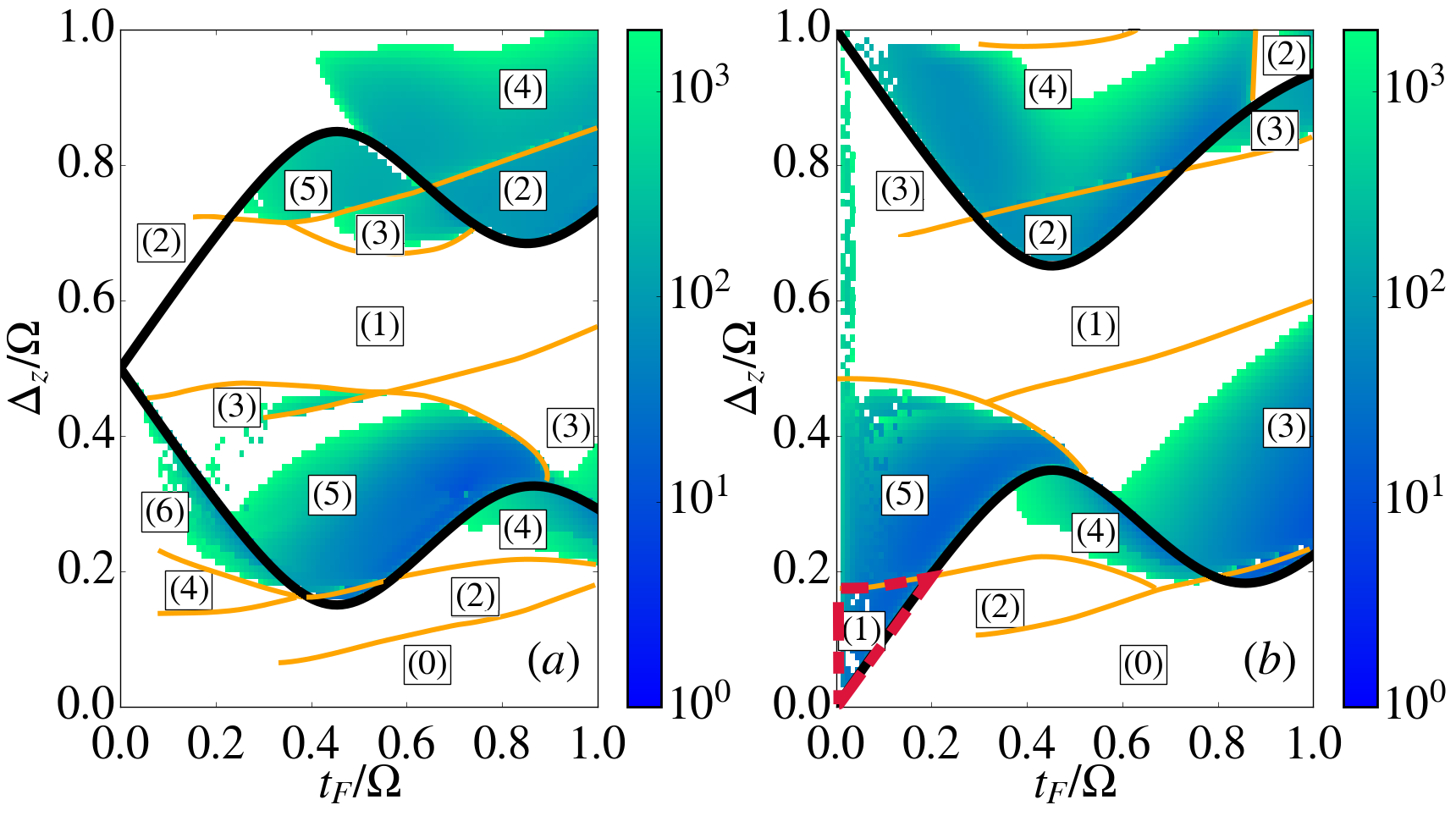}
 \caption{Phase diagrams for the tight-binding Floquet Hamiltonians (a) $h^{e,3}_{nn'}$ (6 replicas) and 
   (b) $h^{o,2}_{nn'}$ (5 replicas) on a half-infinite system (all other parameters as in 
   Fig.~\ref{fig:exdia_tes_65}).}  
 \label{fig:tes_65}
\end{figure}

\section{Topological phase diagram for $3$ Floquet replicas}
\label{sec:3RWA}

In this section we consider the case of $3$ Floquet replicas for the specific model considered in the main
part of the paper \cite{kennes_etal}. In Fig.~\ref{fig:tes_3} we show the topological phase diagram and 
relate it in Fig.~\ref{fig:roots_3RWA} to the evolution of the real parts of the roots of 
$D_k^+=\text{det} A_k^+=0$ along the
dashed line in Fig.~\ref{fig:tes_3}. For higher truncation order the number of roots increases and
the evolution of the roots becomes more complex. However, there are many roots with large imaginary
part (the thin lines in Fig.~\ref{fig:roots_3RWA}(a)) which do not change the sign of their imaginary
part and contribute only weakly to the TESs (if present). Therefore, they do not influence the number $Z$
of TESs and do not participate in the phase transitions. 
The roots with imaginary part below a certain treshold (indicated by thick lines in 
Fig.~\ref{fig:roots_3RWA}) are the most important ones and are related to the local minima
of the bulk band structure (shown in the main part of the paper for ${\Delta_Z\over\Omega}=0.65$). 
As can be seen there are two phase transitions of type ``B'' (orange lines
in Fig.~\ref{fig:tes_3} and orange dashed vertical lines in Fig.~\ref{fig:roots_3RWA}), where
two roots with finite real parts at $\pm k_i$ simultaneously cross the real axis and change the sign
of their imaginary part. According to (\ref{eq:Z}) this changes the number $Z$ of TES by two. In addition,
there are two phase transitions of type ``B'' (black lines in Fig.~\ref{fig:tes_3} and black dashed 
vertical lines in Fig.~\ref{fig:roots_3RWA}), where only one root is crossing through $k=0$ along the
imaginary axis, which changes the number $Z$ of TES by one. 
Both phase transitions of type ``B'' are associated with a bifurcation point lying
very close to the phase transition (shown on a smaller scale in Fig.~\ref{fig:roots_3RWA}(b,c)).

\section{Exact diagonalization study and nontopological edge states}
\label{sec:ntes}

In this section we present the results for the phase diagram of the specific model discussed in the main part
of the paper \cite{kennes_etal} from an exact diagonalization study of
the Hamiltonians $h_{nn'}^{e/o,l_{\text{max}}}$ on a finite system with $N=10000$ unit cells and discuss
the occurrence of nontopological edge states (NTESs) at a finite energy distance from the gap centers.

Using exact diagonalization the results for the localization length of the strongest-localized TES 
are shown in Fig.~\ref{fig:exdia_tes_65}. To identify the TESs and distinguish them
from NTESs we consider only those edge states in this figure which lie very close to the energies 
$E=-{\Omega\over 2}$ and $E=0$, i.e., very deep in the bulk gap. For a finite system, due to parity symmetry 
$U|n\sigma\eta l\rangle=|N-n+1,-\sigma\eta l\rangle$ of the considered model (where $n$, $\sigma$, $\eta$, and $l$ 
denote the indizes for the unit cells, spin up/down, conduction/valence band,
and Floquet modes, respectively), each TES will occur as a pair with parity $\pm$. The energy will not
be exactly in the gap center since two edge states located only at the left/right end of the wire will have a 
small overlap leading to a small splitting in symmetric/antisymmetric TESs w.r.t. parity. This splitting
is exponentially small and, therefore, the TESs are found in the exact diagonalization study from the states
lying very close to the gap centers at $E=-{\Omega\over 2}$ or $E=0$. Taking the criterium that the energy
should lie within $15\%$ of the bulk gap, the phase diagram for the localization length of the
strongest-localized TES are roughly consistent with the results from the half-infinite wire shown in 
Fig.~\ref{fig:tes_65}. There is still some noise visible in the exact diagonalization study since the
criterium to find the TES deep in the bulk gap requires a precise determination of the bulk gap. This is
often quite difficult since the bulk gap is dominated by very small gaps from anticrossings at large quasimomentum. 
In addition, one can also determine the number of TESs in each phase region from the exact diagonalization 
(not shown in Fig.~\ref{fig:exdia_tes_65}) and finds a good agreement with Fig.~\ref{fig:tes_65} . 
Only for small $t_F$ numerical instabilities lead to very strong noise and a comparison is not possible.

\begin{figure}
\centering
 \includegraphics[height=5cm]{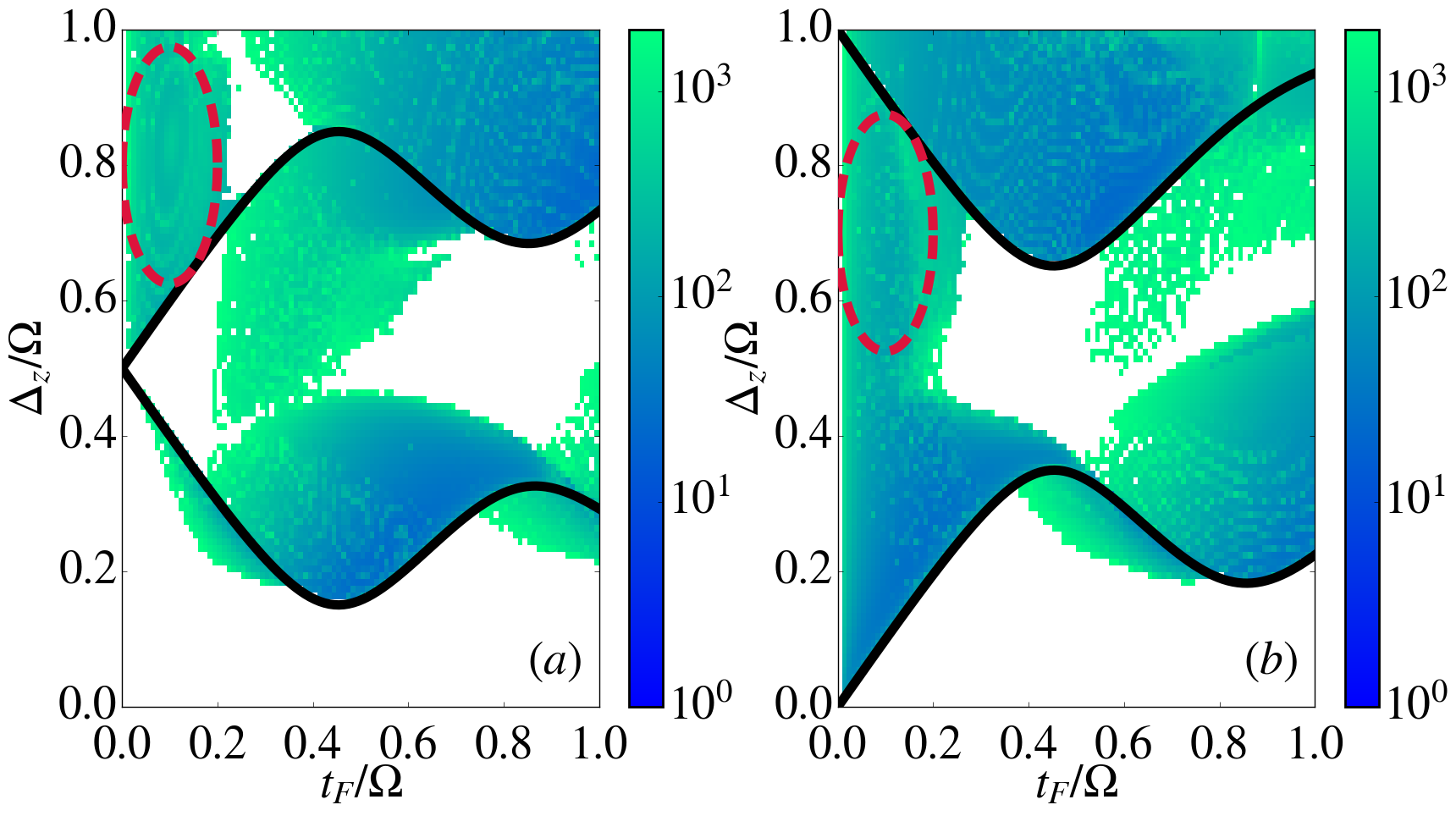}
 \caption{Exact diagonalization study as in Fig.~\ref{fig:exdia_tes_65} but 
   showing the strongest-localized edge state out of the first $20$ eigenstates lying closest to
   the energy (a) $E=-{\Omega\over 2}$ and (b) $E=0$. The regions where additional NTESs show up
   with very short localization length are indicated by red-dashed contours.}
\label{fig:ntes_65}
\end{figure}
To reveal NTESs at finite energy (either located in the gap or in the bulk spectrum),
we have shown in Fig.~\ref{fig:ntes_65} the localization length
of the strongest-localized edge state out of the first $20$ eigenstates lying closest to the center of the gap
(for the same number of Floquet replicas as in Fig.~\ref{fig:tes_65}).  
In comparison to Fig.~\ref{fig:tes_65} additional NTESs appear for $\Delta_Z\gtrsim {\Omega\over 2}$. Of particular
interest are the ones with very short localization length and energies within the bulk gap which are
indicated by a red dashed contour. These edge states are nondegenerate and can not stay in the bulk gap in the 
thermodynamic limit $N\rightarrow\infty$ since they have definite parity (only two degenerate edge states 
with different parities can be localized at the left or right end of the wire by a linear combination).
They occur since the bulk gap is of $O(1/N)$ for small $t_F$ determined from anticrossings with
large quasimomentum. The bulk states at this anticrossing are discrete states with level spacing of the
order of the gap. The NTES is formed by a linear combination of an edge state from anticrossings with
small quasimomentum and large gap (leading to a short localization length) with one of the discrete bulk states
of the anticrossings with large quasimomentum and small gap. This pushes the NTES slightly back into the bulk gap. 
However, for $N\rightarrow\infty$, the NTES will not repel from a discrete bulk state but will be smeared out
via a coupling to the continuum of
states from the anticrossings at large quasimomentum with very small gap. This gives the NTES a small broadening 
and its quasienergy will lie in the bulk. This is similiar to broadening effects of TESs when anticrossings with small
gap at large quasimomentum are closed by inelastic interactions. We note that, for small $t_F$, the NTESs indicated
in the red-dashed contour of Fig.~\ref{fig:ntes_65} have even a smaller localization length compared to the TESs
occurring in these phase regions. Therefore, they will also be observable in STM experiments but their energy is
very unstable such that they can not be used for quantum computing operations. 

In contrast, all other NTESs shown in Fig.~\ref{fig:ntes_65} (not within the dashed contours) are NTESs 
lying already in the bulk spectrum. They have a very strong contribution from bulk states such that only a
fraction of the particle is located at the edge. Nevertheless, the localization length calculated from the
inverse participation ratio is rather small since both the edge and the bulk state forming a linear 
combination for the NTES are normalized wave functions leading to a very small prefactor $\sim 1/\sqrt{N}$ for the 
bulk contribution.